\documentclass[preprint,showpacs,preprintnumbers,amsmath,amssymb]{revtex4}

\usepackage{graphicx}

\begin{document}

\title{
Microlensed image centroid motions 
by an exotic lens object with 
negative convergence or negative mass 
}
\author{Takao Kitamura}
\author{Koji Izumi}
\author{Koki Nakajima}
\author{Chisaki Hagiwara}
\author{Hideki Asada} 
\affiliation{
Faculty of Science and Technology, Hirosaki University,
Hirosaki 036-8561, Japan} 

\date{\today}

\begin{abstract}
Gravitational lens models with negative convergence  
inspired by modified gravity theories, exotic matter and energy 
have been recently examined, 
in such a way that a static and spherically symmetric modified 
spacetime metric depends on the inverse distance to 
the $n$-th power 
($n=1$ for Schwarzschild metric, $n=2$ for Ellis wormhole, 
and $n \neq 1$ for an extended spherical distribution of matter 
such as an isothermal sphere) 
in the weak-field approximation 
[Kitamura, Nakajima and Asada, PRD 87, 027501 (2013), 
Izumi et al. PRD 88 024049 (2013)]. 
Some of the models 
act as if  
a convex lens, 
whereas the others are repulsive on light rays 
like a concave lens. 
The present paper considers microlensed image centroid motions 
by the exotic lens models. 
Numerical calculations show that, for large $n$ cases 
in the convex-type models, 
the centroid shift from the source position 
might move on a multiply-connected curve 
like a bow tie, while it is known to move on an ellipse 
for $n=1$ case and to move on an oval curve for $n=2$. 
The distinctive feature of the microlensed image centroid may be used 
for searching (or constraining) localized exotic matter or energy 
with astrometric observations. 
It is shown also that the centroid shift trajectory 
for concave-type repulsive models 
might be elongated vertically to the source motion direction 
like a prolate spheroid, whereas that for convex-type models 
such as the Schwarzschild one 
is tangentially elongated like an oblate spheroid. 
\end{abstract}


\maketitle

\section{Introduction}
The bending of light 
is among the first experimental confirmations 
of the theory of general relativity. 
As one of the important tools in modern astronomy and cosmology, 
the gravitational lensing is widely used 
for investigating extrasolar planets, dark matter and dark energy. 

The light bending is also of theoretical importance, 
in particular for studying a null structure of a spacetime. 
A rigorous form of the bending angle plays an important role 
in understanding properly a strong gravitational field  
\cite{Frittelli, VE2000, Virbhadra, VNC, VE2002, VK2008, ERT, Perlick}. 
For example, 
strong gravitational lensing in a Schwarzschild black hole 
was considered by Frittelli, Kling and Newman \cite{Frittelli},  
by Virbhadra and Ellis \cite{VE2000} 
and more comprehensively by Virbhadra \cite{Virbhadra}; 
Virbhadra, Narasimha and Chitre \cite{VNC}
studied distinctive lensing features of naked singularities. 
Virbhadra and Ellis \cite{VE2002} 
and Virbhadra and Keeton \cite{VK2008} 
later described 
the strong gravitational lensing by naked singularities; 
DeAndrea and Alexander \cite{DA} discussed the lensing 
by naked singularities to test the cosmic censorship  hypothesis;
Eiroa, Romero and Torres \cite{ERT} treated 
Reissner-Nordstr\"om black hole lensing; 
Perlick \cite{Perlick} discussed the lensing 
by a Barriola-Vilenkin monopole 
and also that by an Ellis wormhole. 
 
One of peculiar features of general relativity is that 
the theory admits a nontrivial topology of a spacetime, 
for instance a wormhole. 
An Ellis wormhole is a particular example of the Morris-Thorne 
traversable wormhole class \cite{Ellis, Morris1, Morris2}.
Furthermore, wormholes are inevitably related with 
violations of some energy conditions in physics \cite{Visser}. 
For instance, dark energy is introduced to explain 
the observed accelerated expansion of the universe 
by means of an additional energy-momentum component 
in the right-hand side of the Einstein equation. 
Furthermore, the left-hand side of the Einstein equation, 
equivalently the Einstein-Hilbert action, 
could be modified in various ways (nonlinear curvature terms, 
higher dimensions, and so on) 
inspired by string theory, loop quantum gravity and so on. 
Because of the nonlinear nature of gravity, 
modifications to one (or both) side of the Einstein equation 
might admit spacetimes significantly different from 
the standard Schwarzschild spacetime metric, 
even if the spacetime is assumed to be asymptotically flat, 
static and spherically symmetric. 
One example is an Ellis wormhole (being an example of 
traversable wormholes). 

Many years ago, 
scattering problems in wormhole spacetimes were discussed 
(for instance, \cite{CC, Clement}). 
Interestingly, the Ellis wormhole has a zero mass 
at the spatial infinity but it causes the light deflection 
\cite{CC, Clement}. 
Moreover, the gravitational lensing by wormholes has been recently 
investigated as an observational probe of such an exotic spacetime 
\cite{Safonova, Shatskii, Perlick, Nandi, Abe, Toki, Tsukamoto, 
Tsukamoto2, Yoo}. 
Several forms of the deflection angle by the Ellis wormhole 
have been recently derived and often used 
\cite{Perlick, Nandi, DS, BP, Abe, Toki, Tsukamoto}. 
A reason for such differences has been clarified by 
several authors \cite{Nakajima, Gibbons}. 

Small changes in gravitational lensing 
in modified gravity theories such as $f(R)$ and fourth-order gravity 
have been studied 
(e.g. \cite{Capozziello, Horvath, Mendoza, Asada2011}). 
Furthermore, Horvath, Gergely, and Hobill \cite{HGH} 
studied lensing effects 
with negative convergence 
by so-called tidal charges in the Dadhich et al. solution, 
where, for a brane world black hole, the tidal charge 
arises from tidal forces acting at the brane-bulk boundary 
\cite{Dadhich}. 
A point is that negative convergence in this case does not require 
any exotic matter. 
It comes from the Weyl curvature in higher dimensions.

Inspired by a number of works on modifications in gravitational lensing, 
Kitamura et al. \cite{Kitamura} assume, in a phenomenological sense, 
that an asymptotically flat, static and spherically symmetric 
modified spacetime could depend on 
the inverse distance to the power of positive $n$ 
in the weak field approximation. 
The Schwarzschild spacetime and the Ellis wormhole correspond to 
$n=1$ and $n=2$, respectively, so that these spacetimes 
can be expressed as a one-parameter family. 
This one-parameter model expresses a spherical mass distribution. 
Note that Birkhoff's theorem could say that cases $n \neq 1$ 
might be non-vacuum,  
if the models were interpreted in the framework of 
the standard Einstein equation.

Kitamura et al. \cite{Kitamura} 
have shown that demagnification could occur for $n>1$ including 
the Ellis wormhole case ($n=2$). 
They have also shown that time-symmetric demagnification parts 
might appear in light curves due to gravitational microlensing effects 
by such exotic models, where light curves are useful 
in microlensing observations in our galaxy. 
For cosmological situations, however, the Einstein ring size 
becomes so large and hence the typical time scale is so long 
that light curves cannot be observable in cosmology. 
On the other hand, the image separation angle becomes sufficiently large, 
so that it can be practically measured. 
By using the latest result in the Sloan Digital Sky Survey Quasar Lens 
Search, Takahashi and Asada have recently set the first upper bound 
on the cosmic abundances of Ellis wormholes and also 
negative-mass compact objects \cite{Takahashi}. 
In theoretical physics, negative mass is a hypothetical concept 
of matter whose mass is of opposite sign to the mass of normal matter. 
Although possible negative mass ideas have been often discussed 
since the 19th century, there has been no evidence for them 
\cite{Bondi,Jammer1961,Jammer1999,Cramer}. 
The negative masses might attract each other 
to form a negative massive clump, so that 
such clumps could reside in cosmological voids (e.g. \cite{Piran}). 
Gibbons and Kodama \cite{GK} have shown that curvature-regular 
asymptotically flat solitons with negative mass are 
contained in the Myers-Perry family, 
though the soliton solutions in the odd spacetime dimensions 
might not express real astrophysical objects.

However, the information on the image separation angle 
is not sufficient for distinguishing exotic lens models. 
Hence, Izumi et al. \cite{Izumi} have investigated 
gravitational lensing shear by an exotic lens object with 
negative convergence or negative mass. 
They have shown that 
images by the lens models for the gravitational pull 
(like a convex lens in optics) are tangentially elongated, 
whereas those by the repulsive ones (like a concave lens) 
are radially distorted. 
Their study \cite{Izumi} might concern the strong (or weak) 
lensing surveys at the extra-galactic or cosmological distance. 

Therefore, the main purpose of the present paper 
is to study microlensed image centroid motions 
by such exotic gravitational lens models.  
Here, we focus on the microlensing in our galaxy. 
Studies of centroid displacements of lensed images 
have been often done for the Schwarzschild lens 
\cite{Walker,MY,HOF,SDR,JHP,Lewis,asada02,HL}. 
Virbhadra and Keeton \cite{VK2008} have investigated 
the centroid displacement for naked singularities by using 
the Janis-Newman-Winicour solution. 
Toki et al. \cite{Toki} have studied the centroid motion 
by Ellis wormhole.  
The main results of the present paper are: 
(1) For certain exotic lens models,  
the centroid shift from the source position 
might move on a multiply-connected curve 
like a {\it bow tie} for large $n$ cases, 
while it is known to move on 
an ellipse for $n=1$ case \cite{Walker,JHP} 
and to move on an oval curve for $n=2$ \cite{Toki}. 
(2)For concave-type repulsive lens models, 
the centroid displacement might move on a simply-connected curve 
but might be elongated vertically to the source velocity, 
while it is tangentially elongated for Schwarzschild case. 

We take the units of $G=c=1$ throughout this paper.

\section{Modified spacetime model and modified lens equation}
This section briefly summarizes the basics 
of the exotic lens models \cite{Kitamura,Izumi}. 
\subsection{Modified bending angle of light}
Following Kitamura et al. \cite{Kitamura},  
the present paper assumes that 
an asymptotically flat, static and spherically symmetric 
modified spacetime could depend on 
the inverse distance to the power of positive $n$ 
in the weak field approximation. 
We consider the light propagation through a four-dimensional spacetime, 
though the whole spacetime may be higher dimensional. 
The four-dimensional spacetime metric is expressed as 
\begin{equation}
ds^2=-\left(1-\frac{\varepsilon_1}{r^n}\right)dt^2
+\left(1+\frac{\varepsilon_2}{r^n}\right)dr^2
+r^2(d\Theta^2+\sin^2\Theta d\phi^2) 
+O(\varepsilon_1^2, \varepsilon_2^2, \varepsilon_1 \varepsilon_2) ,  
\label{ds}
\end{equation}
where $r$ is the circumference radius and 
$\varepsilon_1$ and $\varepsilon_2$ are small book-keeping 
parameters in iterative calculations. 
The weak field approximation means  
$\varepsilon_1 / r^n \ll 1$ 
and 
$\varepsilon_2 / r^n \ll 1$. 
Namely, we study a far field from the lens object as 
$r \gg \varepsilon_1^{1/n}$ and $r \gg \varepsilon_2^{1/n}$. 
Note that Eq. (\ref{ds}) is not valid 
in the strong field near $r=0$ 
(Please see \cite{Note-1} for more detail). 
Here, $\varepsilon_1$ and $\varepsilon_2$ 
may be either positive or negative, respectively. 
Negative $\varepsilon_1$ and $\varepsilon_2$ for $n=1$ 
correspond to a negative mass (in the linearized Schwarzschild metric). 

Without loss of generality, we focus on 
the equatorial plane $\Theta = \pi/2$, 
since the spacetime is spherically symmetric. 
The deflection angle of light 
is obtained at the linear order as \cite{Kitamura} 
\begin{align}
\alpha
&=\dfrac{\varepsilon}{b^n}\int_0^{\frac{\pi}{2}} \cos^n\psi d\psi 
+O(\varepsilon^2) , 
\label{alpha}
\end{align}
where the integral is positive definite, 
$b$ denotes the impact parameter of the light ray, 
and we define $\varepsilon \equiv n \varepsilon_1 + \varepsilon_2$. 
By absorbing the positive integral 
into the parameter $\varepsilon$, we rewrite the linear-order 
deflection angle simply as 
$\alpha = \bar\varepsilon/b^n$, 
where the sign of $\bar\varepsilon$ is the same as that of $\varepsilon$. 
This deflection angle recovers 
the Schwarzschild ($n=1$) and Ellis wormhole ($n=2$) cases. 
For $\varepsilon > 0$, the deflection angle of light is always positive, 
which means that the corresponding spacetime model causes 
the gravitational pull on light rays. 
For $\varepsilon < 0$, on the other hand, it is inevitably negative, 
which implies the gravitational repulsion on light rays 
like a concave lens.

We mention an effective mass. 
A simple application of the standard lens theory \cite{SEF} 
suggests that the deflection angle of light in the form of 
$\alpha = \bar\varepsilon/b^n$ 
corresponds to a convergence (scaled surface mass density) as 
\begin{equation}
\kappa(b) = \frac{\bar\varepsilon (1-n)}{2} \frac{1}{b^{n+1}} , 
\label{kappa}
\end{equation}
which implies an extended spherical distribution of matter (or energy)  
for $n \neq 1$ and a singular source only for $n = 1$.

For the weak-field Schwarzschild case ($n = 1$), 
it follows that the convergence vanishes. 
For $\varepsilon > 0$ and $n>1$, 
the effective surface mass density of the lens object 
is interpreted as negative in the framework 
of the standard lens theory \cite{Kitamura}. 
This means that the matter (and energy) needs to be exotic 
if $\varepsilon > 0$ and $n>1$. 
Also when $\varepsilon < 0$ and $n<1$, 
the convergence is negative and hence 
the matter (and energy) needs to be exotic. 
Interestingly, when $\varepsilon < 0$ and $n>1$, 
the convergence is positive everywhere except for the central singularity 
and hence exotic matter (and energy) is not required in the framework 
of the standard lens theory, in spite of the gravitational repulsion on 
light rays. 
Attraction ($\varepsilon > 0$) and repulsion ($\varepsilon < 0$) 
in the above two-parameter models do not have a one-to-one correspondence 
to positive convergence $\kappa > 0$ and negative one $\kappa < 0$. 
This point is summarized in Table \ref{table-1} \cite{Izumi}.

\subsection{Modified Einstein radius}
Under the thin lens approximation, 
it is useful to consider the lens equation as \cite{SEF} 
\begin{equation}
\beta = \frac{b}{D_{\it{L}}} - \frac{D_{\it{LS}}}{D_{\it{S}}} \alpha(b) , 
\label{lenseq}
\end{equation}
where 
$\beta$ denotes the angular position of the source and 
$D_{\it{L}}$, $D_{\it{S}}$, $D_{\it{LS}}$ are the distances from the observer 
to the lens, from the observer to the source, and from the lens to
the source, respectively. 
Note that there is the mathematical consistency of the use 
of the lens equation Eq. (\ref{lenseq}), 
where the trigonometric functions are approximated by 
their leading terms. 
The present paper studies the leading term in the deflection angle, 
so that Eq. (\ref{lenseq}) can be mathematically consistent. 
On the other hand, if one wishes to include the next (and higher order) 
for the bending angle, the third-order (or higher-order) terms 
in the expansion of the trigonometric functions have to be 
taken into account in the lens equation, 
because of the mathematical consistency 
\cite{Perlick,VE2000,VE2002}.

For $\varepsilon > 0$, 
there is always a positive root corresponding to 
the Einstein ring for $\beta=0$. 
The Einstein ring radius is defined as \cite{Izumi} 
\begin{equation}
\theta_{\it{E}} \equiv 
\left(
\frac{\bar\varepsilon D_{\it{LS}}}{D_{\rm{S}} D_{\it{L}}^n}
\right)^{\frac{1}{n+1}} .
\label{theta_E}
\end{equation} 
If $\varepsilon < 0$, on the other hand, 
Eq. (\ref{lenseq}) has no positive root for $\beta = 0$. 
This is because this case describes the repulsive force. 
For later convenience in normalizing the lens equation, 
we define the (tentative) Einstein ring radius for $\varepsilon < 0$ 
as 
\begin{equation}
\theta_{\it{E}} \equiv 
\left(
\frac{|\bar\varepsilon| D_{\it{LS}}}{D_{\it{S}} D_{\it{L}}^n}
\right)^{\frac{1}{n+1}} , 
\label{theta_E2}
\end{equation} 
though the Einstein ring does not appear for this case. 
This radius gives a typical angular size for $\varepsilon < 0$ lenses. 

Like Schwarzschild lenses, there might exist a photon sphere 
for $\varepsilon > 0$. 
The radius of the photon sphere for the spacetime metric 
by Eq. (\ref{ds}) might become 
\begin{equation}
R_{ps} = \left( \frac{(n+2)\varepsilon_1}{2} \right)^{1/n} .
\end{equation}
See \cite{CVE} for a more thorough discussion on the photon surfaces.

\subsection{Modified lens equation: $\varepsilon > 0$ case}
Following Izumi et al. \cite{Izumi}, 
let us begin with $\varepsilon > 0$ case. 
As already stated, the matter (and energy) needs to be exotic if $n > 1$. 
In the units of the Einstein ring radius, 
Eq. (\ref{lenseq}) is rewritten 
in the vectorial form 
as 
\begin{eqnarray}
\boldsymbol{\hat\beta}
&=& \boldsymbol{\hat\theta} 
- \frac{\boldsymbol{\hat\theta}}{\hat\theta^{n+1}}  
\quad (\hat\theta > 0) , 
\label{lenseqP}\\
\boldsymbol{\hat\beta}
&=& \boldsymbol{\hat\theta} 
- \frac{\boldsymbol{\hat\theta}}{(-\hat\theta)^{n+1}}  
\quad (\hat\theta < 0) , 
\label{lenseqM}
\end{eqnarray}
where we normalize $\hat\beta \equiv \beta/\theta_{\it{E}}$ and 
$\hat\theta \equiv \theta/\theta_{\it{E}}$ 
for the angular position of the image $\theta \equiv b/D_{\it{L}}$,  
and $\boldsymbol{\hat\beta}$ and $\boldsymbol{\hat\theta}$ 
denote the corresponding vectors. 
There is always one image for $\hat\theta > 0$, 
while the other image appears for $\hat\theta < 0$ \cite{Kitamura}.

\subsection{Modified lens equation: $\varepsilon < 0$ case}
\begin{figure}
\includegraphics[width=10cm]{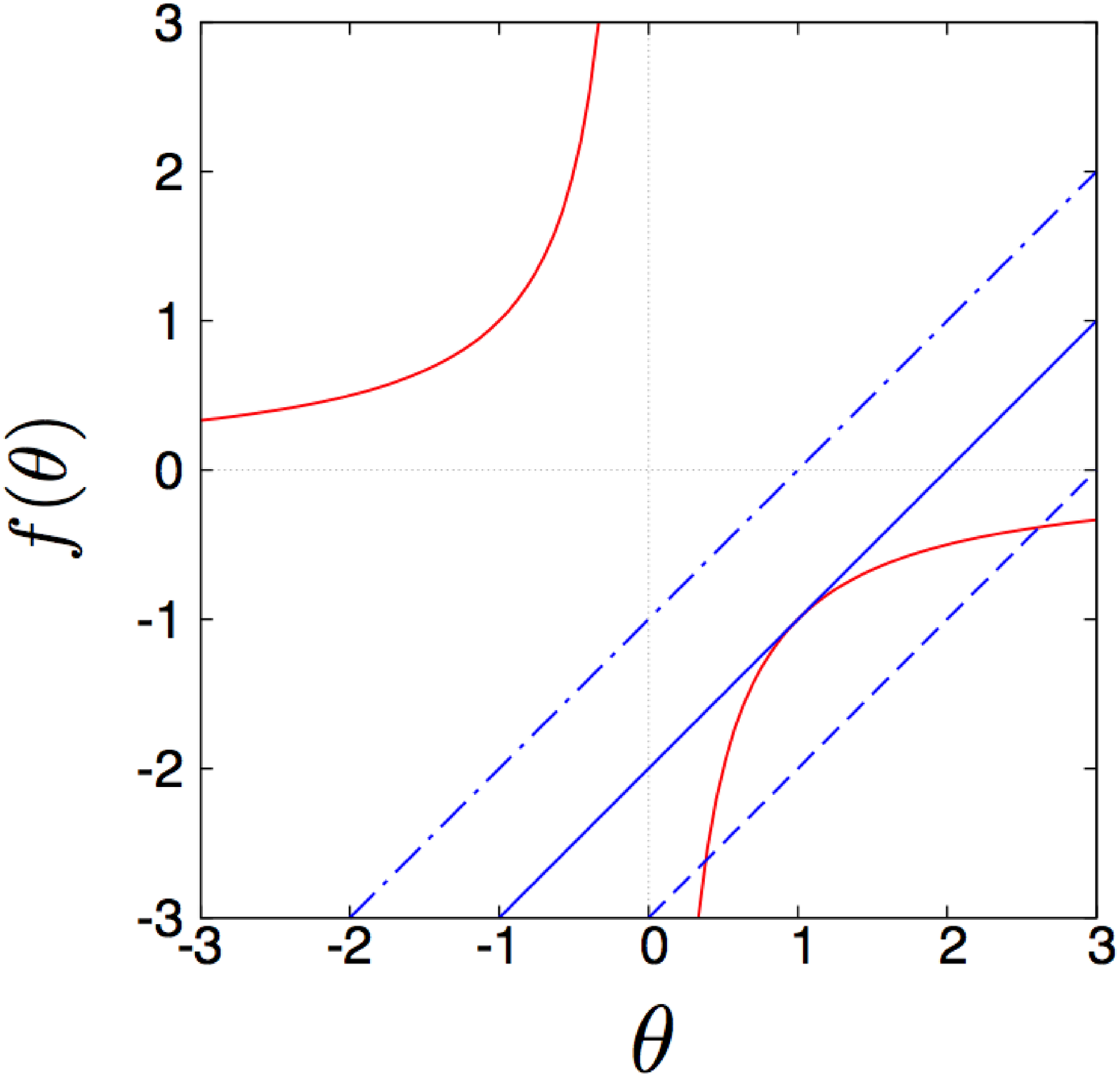}
\caption{
Repulsive lens model $(\varepsilon < 0)$. 
Solid curves denote $1/\hat\theta^n$ and 
straight lines mean $\hat\theta - \hat\beta$. 
Their intersections correspond to image positions 
that are roots for the lens equation. 
There are three cases: No image for a small $\hat\beta$ 
(dot-dashed line), 
a single image for a particular $\hat\beta$ 
(dotted line), 
and two images for a large $\hat\beta$ 
(dashed line). 
The two images are on the same side of the lens object. 
}
\label{figure-1}
\end{figure}

Next, let us mention $\varepsilon < 0$ case \cite{Izumi}. 
In the units of the Einstein ring radius, 
Eq. (\ref{lenseq}) is rewritten 
in the vectorial form 
as 
\begin{eqnarray}
\boldsymbol{\hat\beta}
&=& \boldsymbol{\hat\theta} 
+ \frac{\boldsymbol{\hat\theta}}{\hat\theta^{n+1}}  
\quad (\hat\theta > 0) , 
\label{lenseqP2}\\
\boldsymbol{\hat\beta}
&=& \boldsymbol{\hat\theta} 
+ \frac{\boldsymbol{\hat\theta}}{(-\hat\theta)^{n+1}}  
\quad (\hat\theta < 0) . 
\label{lenseqM2}
\end{eqnarray} 
Without loss of generality, we assume $\hat\beta >0$. 
Then, Eq. (\ref{lenseqM2}) has no root satisfying $\hat\theta < 0$, while 
Eq. (\ref{lenseqP2}) has at most two positive roots. 
Figure \ref{figure-1} shows that there are three cases of the image
number. 
For a large impact parameter case, two images appear on the same side 
with respect to the lens position, 
while no image appears for a small impact parameter. 
The only one image appears only when the impact parameter takes 
a critical value. 
Let us focus on the two image cases, from which the single image case 
can be discussed in the limit as the impact parameter 
approaches the particular value.

\section{Microlensed image centroid} 
\subsection{Image centroid}
Let us study the microlensed image centroid motions. 
In any case of $\varepsilon > 0$ and $\varepsilon < 0$, 
the image positions are denoted by $\boldsymbol{\hat{\theta}}_1$ 
and $\boldsymbol{\hat{\theta}}_2$, and 
the corresponding amplification factors 
are denoted by $A_1$ and $A_2$. 
Without loss of generality, 
we take $\hat\theta_1 > \hat\theta_2$. 
In analogy with the center of the mass distribution, 
the centroid position of the light distribution of 
a gravitationally microlensed source is given by 
\begin{eqnarray}
\boldsymbol{\hat{\theta}}_{\it{pc}} 
&=& 
\frac{A_{1} \boldsymbol{\hat{\theta}}_{1} 
+ A_{2} \boldsymbol{\hat{\theta}}_{2}}{A_{\it{tot}}} , 
\label{pc}
\end{eqnarray}
where $A_{\it{tot}}$ denotes the total amplification as 
$A_{1}+A_{2}$. 
The corresponding scalar is defined as 
$\hat{\theta}_{\it{pc}} \equiv (A_{1} \hat{\theta}_{1} 
+ A_{2} \hat{\theta}_{2}) A_{\it{tot}}^{-1}$. 
Note that $\hat{\theta}_{\it{pc}}$ is positive, when the centroid is 
located on the same side of the source 
with respect to the lens center. 

The relative displacement of the image centroid 
with respect to the source position is written as 
\begin{equation}
\delta\boldsymbol{\hat{\theta}}_{\it{pc}} 
= 
\boldsymbol{\hat{\theta}}_{\it{pc}} 
- \boldsymbol{\hat{\beta}} .
\label{deltapc}
\end{equation}
Henceforth, this is referred to as the centroid shift. 
The corresponding scalar is defined as 
$\delta\hat{\theta}_{\it{pc}} 
\equiv \hat{\theta}_{\it{pc}} - \hat{\beta}$. 
$\delta\hat{\theta}_{\it{pc}}$ is positive, when 
$\hat{\theta}_{\it{pc}}$ is larger than $\hat{\beta}$. 

By taking account of the relation between the lens and 
source trajectory in the sky, 
the time dependence of $\hat{\beta}$ is written as 
\begin{equation}
\hat{\beta}(t) = \sqrt{\hat{\beta}_0^2 + {(t -t_0)^2/t_{\it{E}}}^2}, 
\label{betat}
\end{equation}
where $\hat{\beta}_0$ is the impact parameter of 
the source trajectory and $t_0$ is the time of closest approach. 
Here, the source is assumed to be in uniform linear motion. 
We choose $t_0 = 0$. 
$t_{\it{E}}$ is the Einstein radius crossing time given by
\begin{equation}
t_{\it{E}} = R_{\it{E}} / v_{\it{T}},  \label{eqn:te}
\end{equation}
where $v_{\it{T}}$ is the transverse velocity of the lens relative to the
source and observer. 
In the following numerical computations, 
time is normalized by the Einstein ring radius crossing time. 

In making numerical figures, we employ $x-y$ coordinates, 
such that the coordinate origin is chosen as the lens center, 
$x$-axis is taken along the direction of the source motion 
and $y$-axis is perpendicular to the source motion.

\subsection{Numerical computations: $\varepsilon > 0$ case} 
\begin{figure}
\includegraphics[width=8cm]{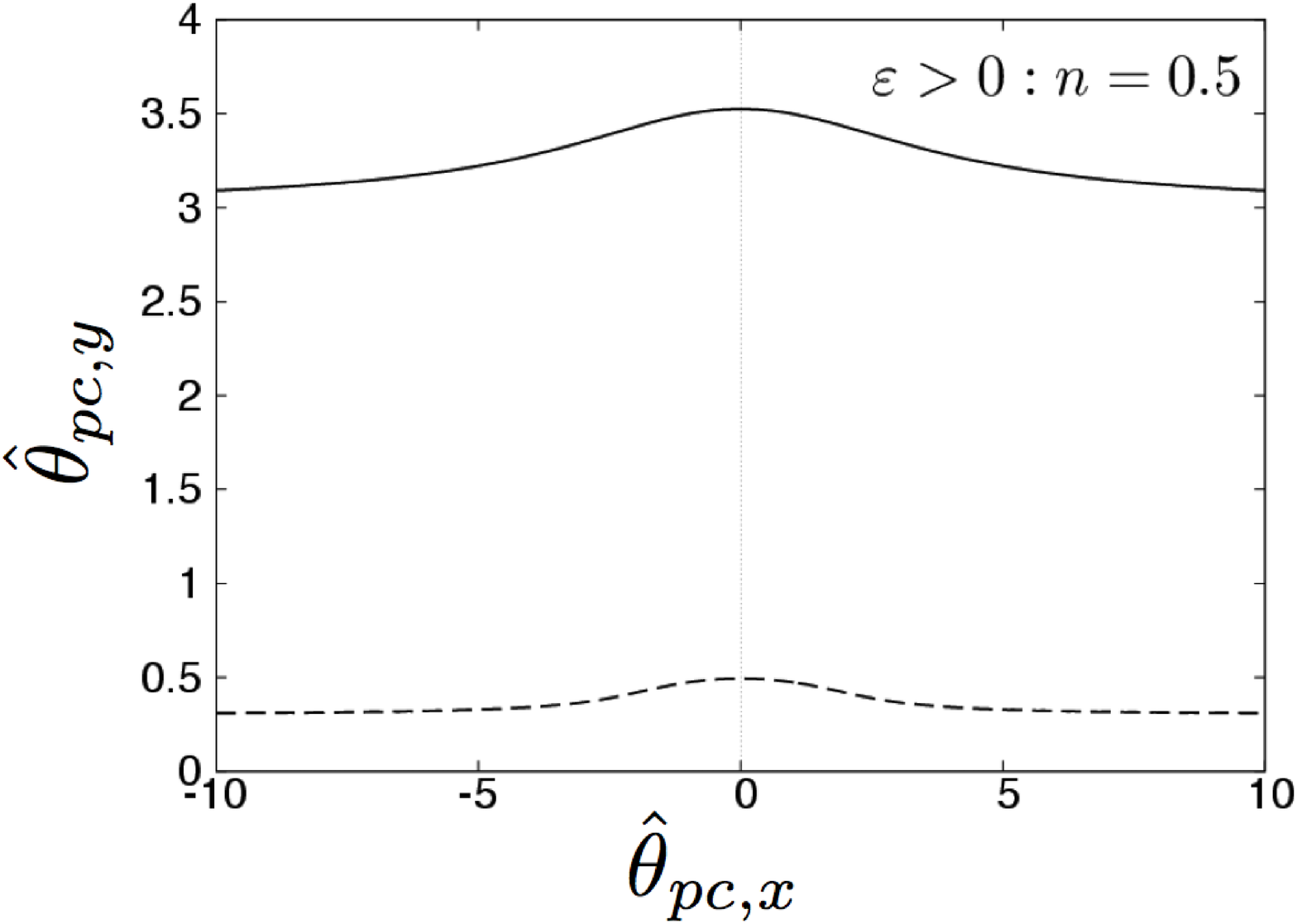}
\includegraphics[width=8cm]{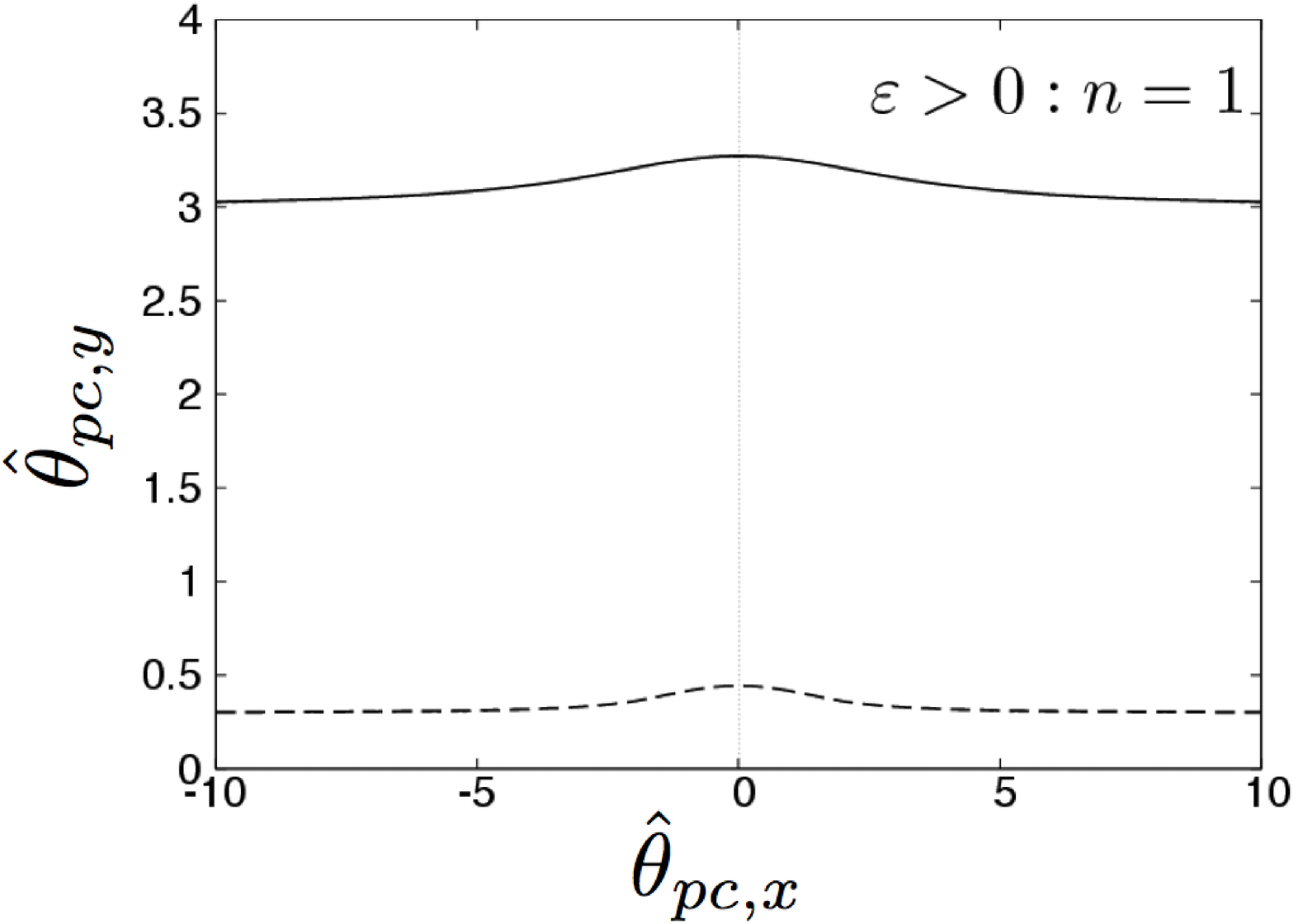}
\includegraphics[width=8cm]{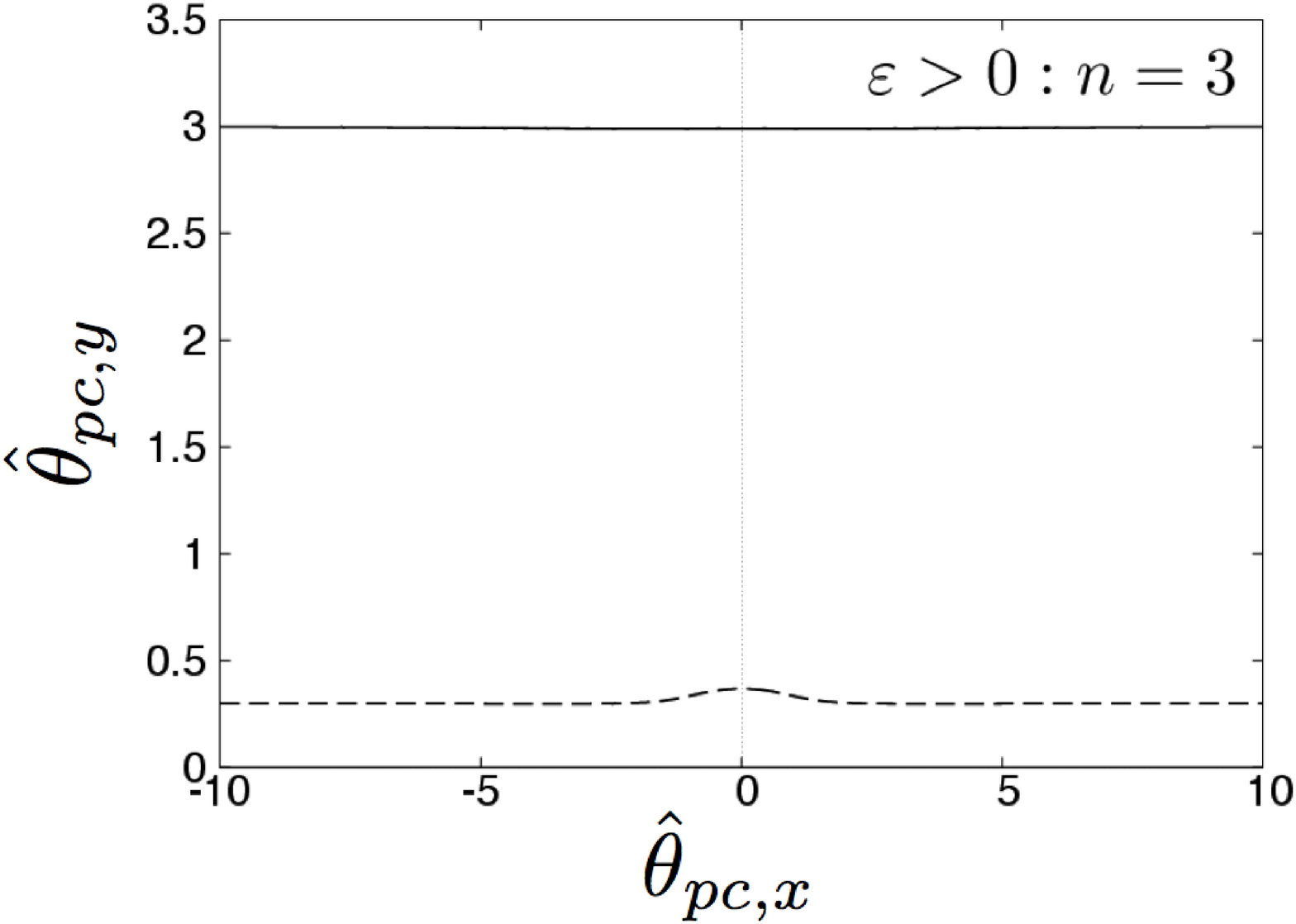}
\includegraphics[width=8cm]{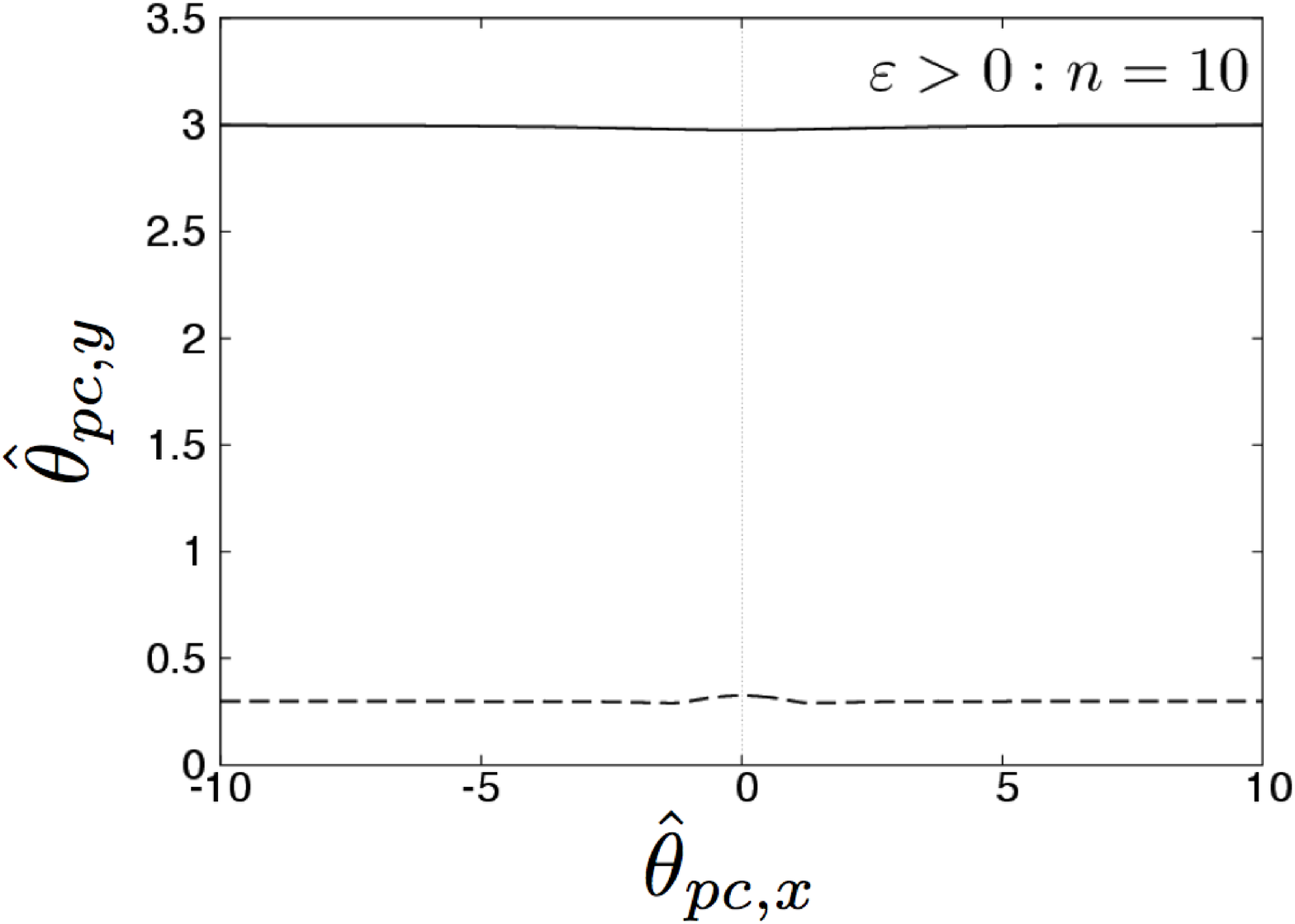}
\caption{
Centroid motions as $(\hat\theta_{\it{pc},x}, \hat\theta_{\it{pc},y})$ 
for $\varepsilon > 0$ (convex-type attractive models). 
The solid and dashed curves 
correspond to $\hat\beta_0 = 3$ and $\hat\beta_0 = 0.3$, respectively. 
The horizontal axis along the source linear motion is $\hat\theta_{\it{pc},x}$ 
and the vertical axis is $\hat\theta_{\it{pc},y}$. 
Top left: $n = 0.5$
Top right: $n = 1$. 
Bottom left: $n = 3$. 
Bottom right: $n = 10$. 
}
\label{figure-2}
\end{figure}

\begin{figure}
\includegraphics[width=8cm]{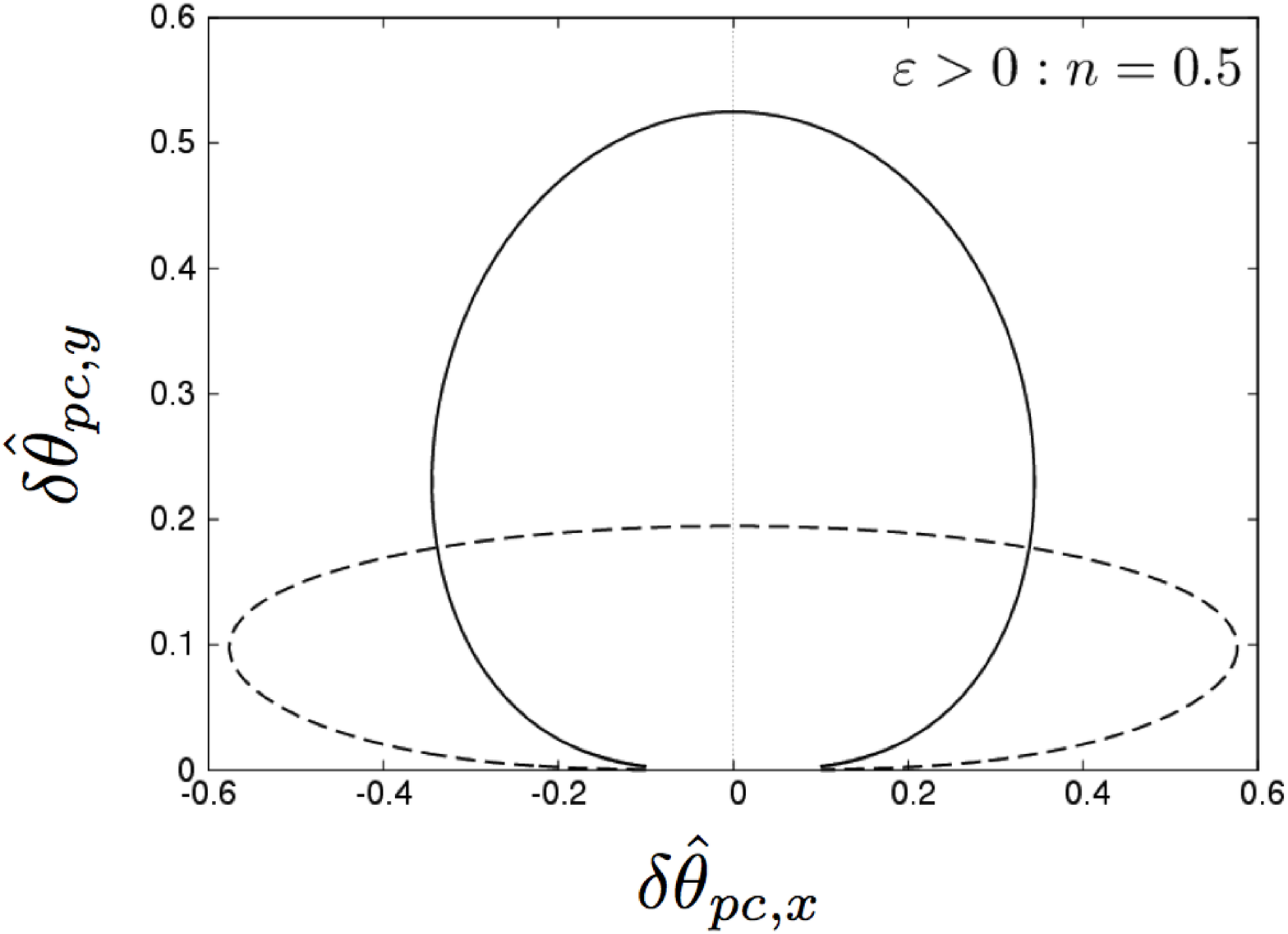}
\includegraphics[width=8cm]{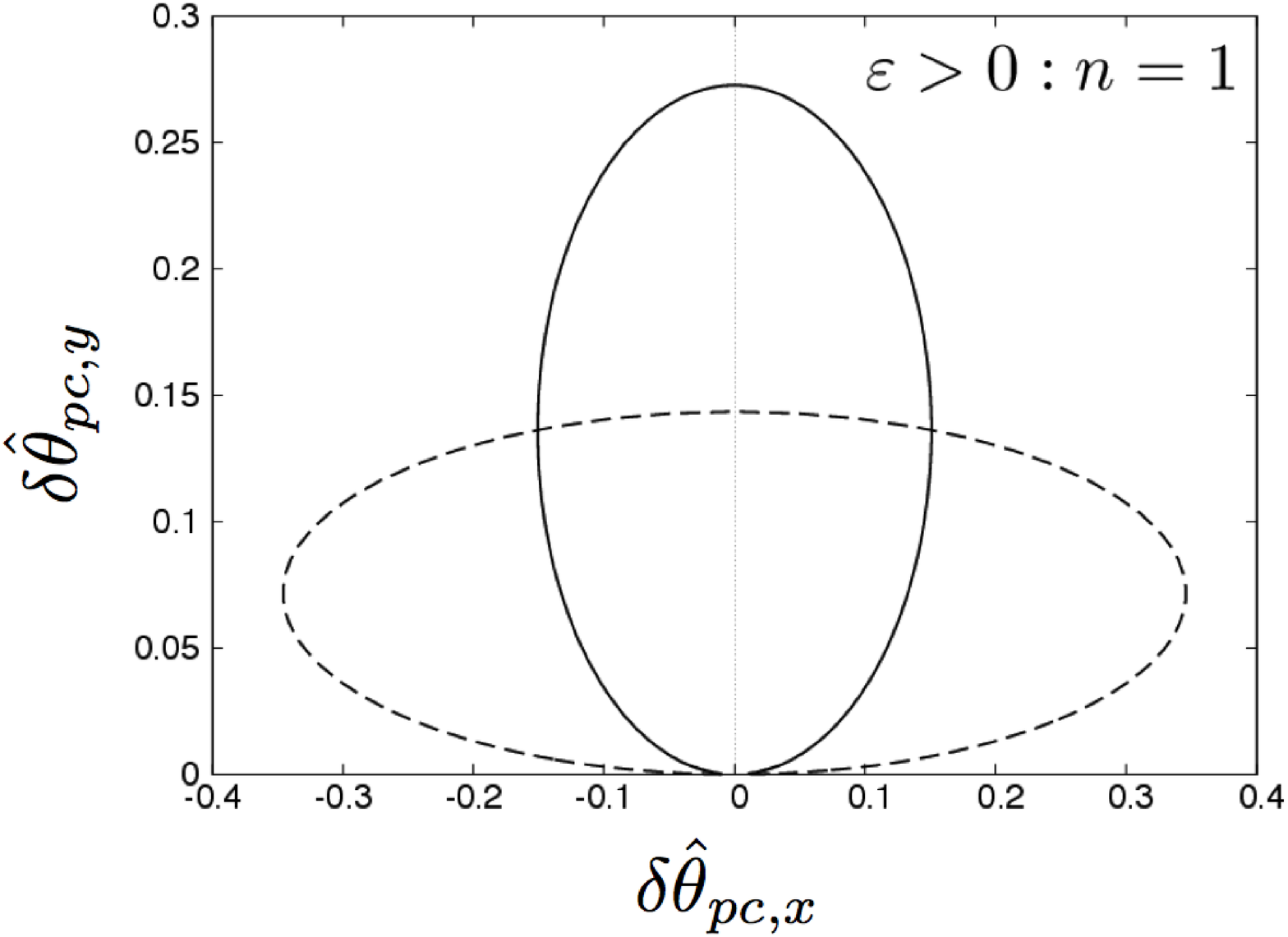}
\includegraphics[width=8cm]{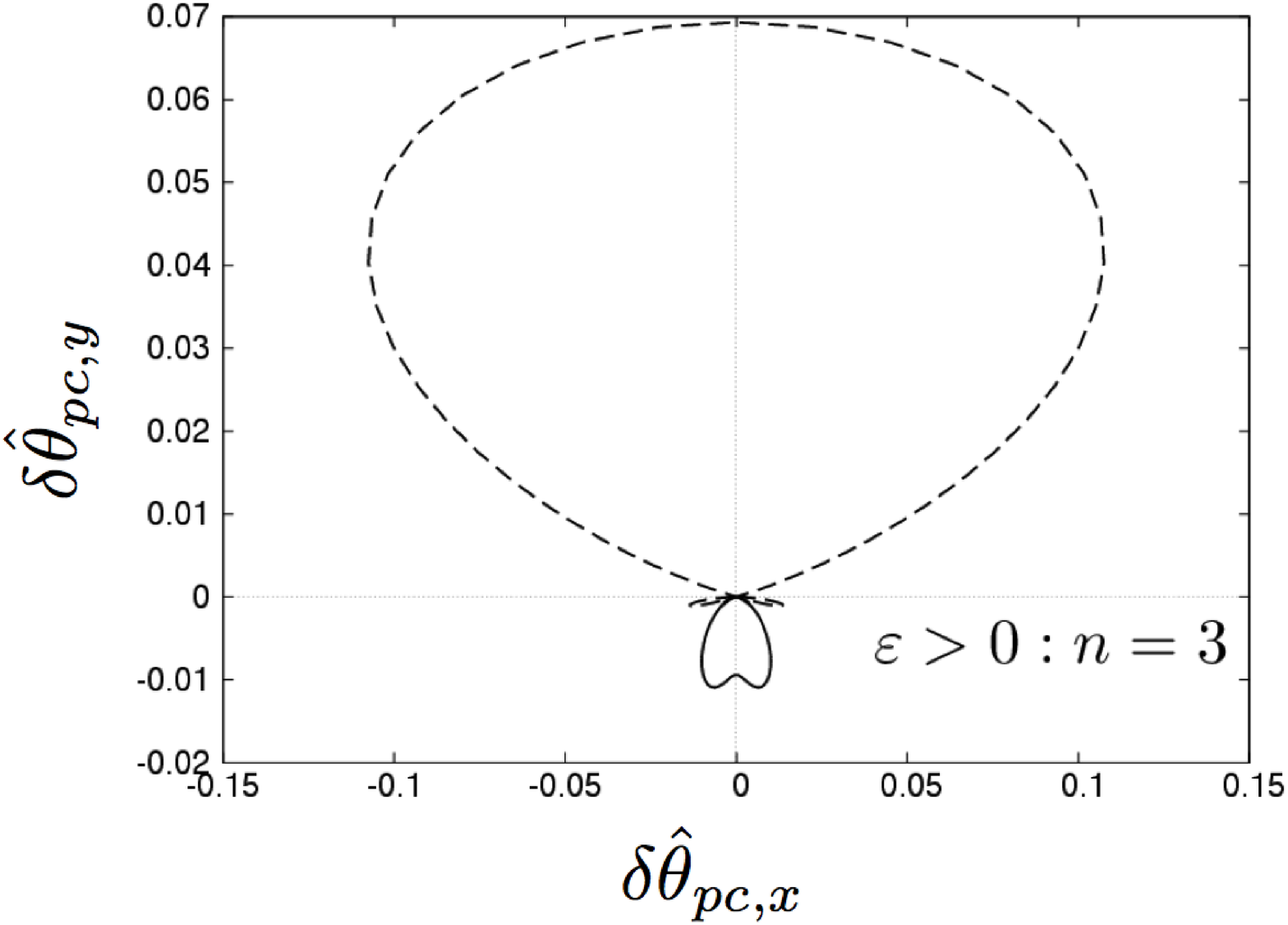}
\includegraphics[width=8cm]{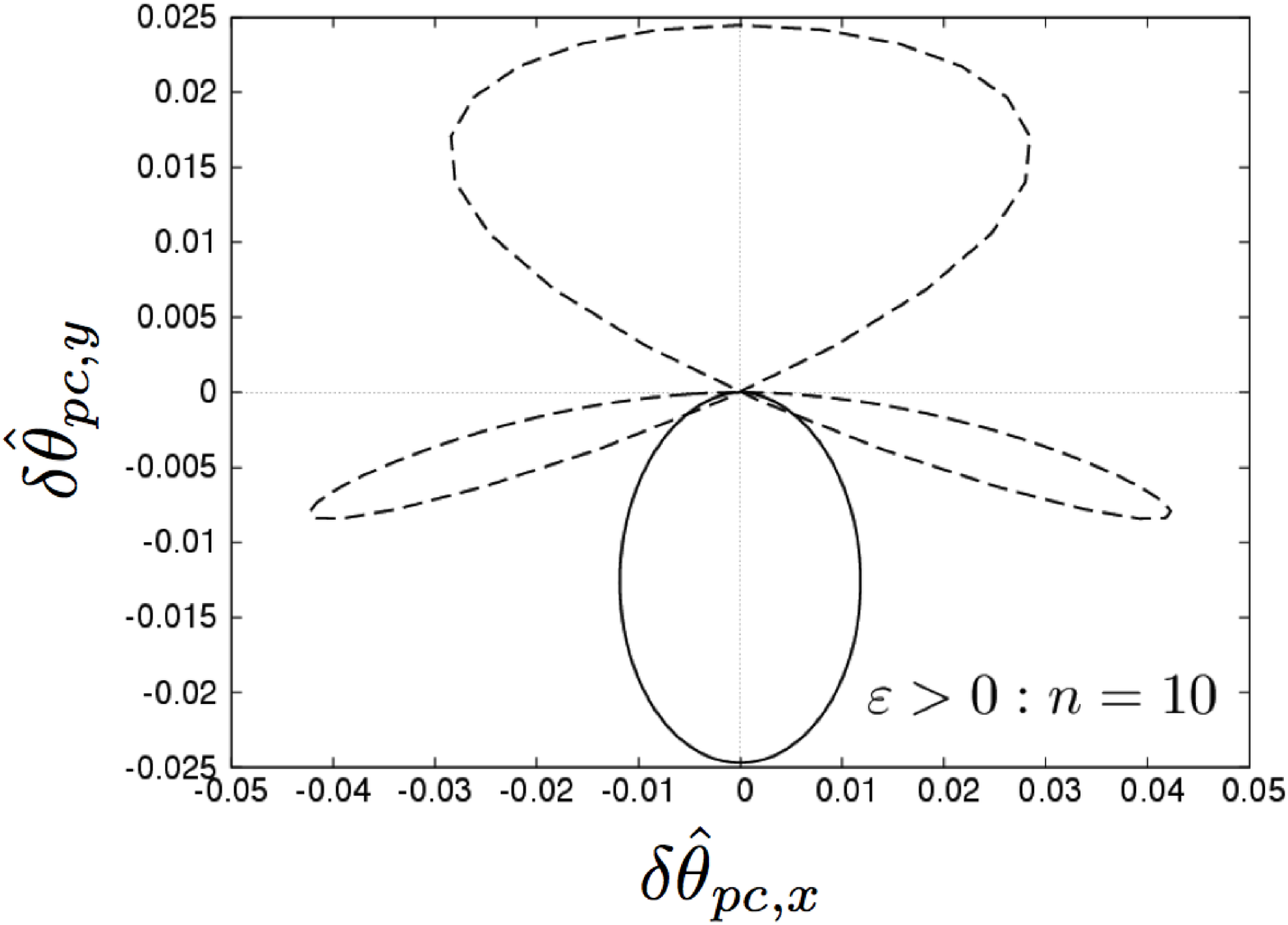}
\caption{
Centroid shifts $\delta\boldsymbol{\hat{\theta}}_{\it{pc}}$ 
for $\varepsilon > 0$ (convex-type attractive models). 
The solid and dashed curves 
correspond to $\hat\beta_0 = 3$ and $\hat\beta_0 = 0.3$, respectively. 
The horizontal axis along the source velocity is $\delta\hat\theta_{\it{pc},x}$ 
and the vertical axis is $\delta\hat\theta_{\it{pc},y}$. 
Top left: $n = 0.5$
Top right: $n = 1$. 
Bottom left: $n = 3$. 
Bottom right: $n = 10$. 
}
\label{figure-3}
\end{figure}

Let us begin with the $\varepsilon > 0$ case. 
See Figure \ref{figure-2} for 
the image centroid trajectories by $\varepsilon > 0$ models 
for $\hat\beta_0 = 0.3$ and $3$. 
Figure \ref{figure-3} shows the image centroid shift 
by the $\varepsilon > 0$ models. 
For $\hat\beta_0 = 0.3$ for instance, 
the maximum vertical shift of the image centroid position 
by the exotic lens models is 
$0.2 (n=0.5)$, $0.14 (n=1)$, $0.07 (n=3)$ and $0.02 (n=10)$ 
in the units of the Einstein ring radius, respectively. 
For $\hat\beta_0 = 3$, 
it is nearly 
$0.5 (n=0.5)$, $0.3 (n=1)$, $-0.01 (n=3)$ and $-0.02 (n=10)$. 
These results suggest that the astrometric lensing by the exotic models 
with large $n$ is relatively weak  
compared with that by the Schwarzschild one ($n=1$). 
In the weak-field region, 
one can understand the suppression of the anomalous shift 
of the image centroid position for large $n$, 
because the bending angle by the large $n$ models  
is proportional to the inverse impact parameter to the power of $n$, 
whereas that by the Schwarzschild lens depends on 
the inverse impact parameter.

A distinctive feature is that 
in $\varepsilon > 0$ and $n > 2$ cases 
bow knots might be added 
into the centroid shift trajectory, 
while the trajectory is known to be an ellipse for $n=1$ case 
\cite{Walker, JHP} and to be oval for $n=2$ \cite{Toki}. 
Such a multiply-connected shape of the centroid shift orbit 
would be an evidence of the corresponding exotic lens 
in astrometric observations. 
Figure \ref{figure-3} shows the bow-tie shape 
might disappear when the impact parameter becomes sufficiently large, 
for instance $\hat\beta \sim 3$. 
For $\varepsilon > 0$ and $n=3$, 
the centroid shift could be negative 
for the $\hat\beta_0 = 3$ case. 
This is partly because 
$A_2$ becomes large compared with the $n=1$ case.

\begin{figure}
\includegraphics[width=8cm]{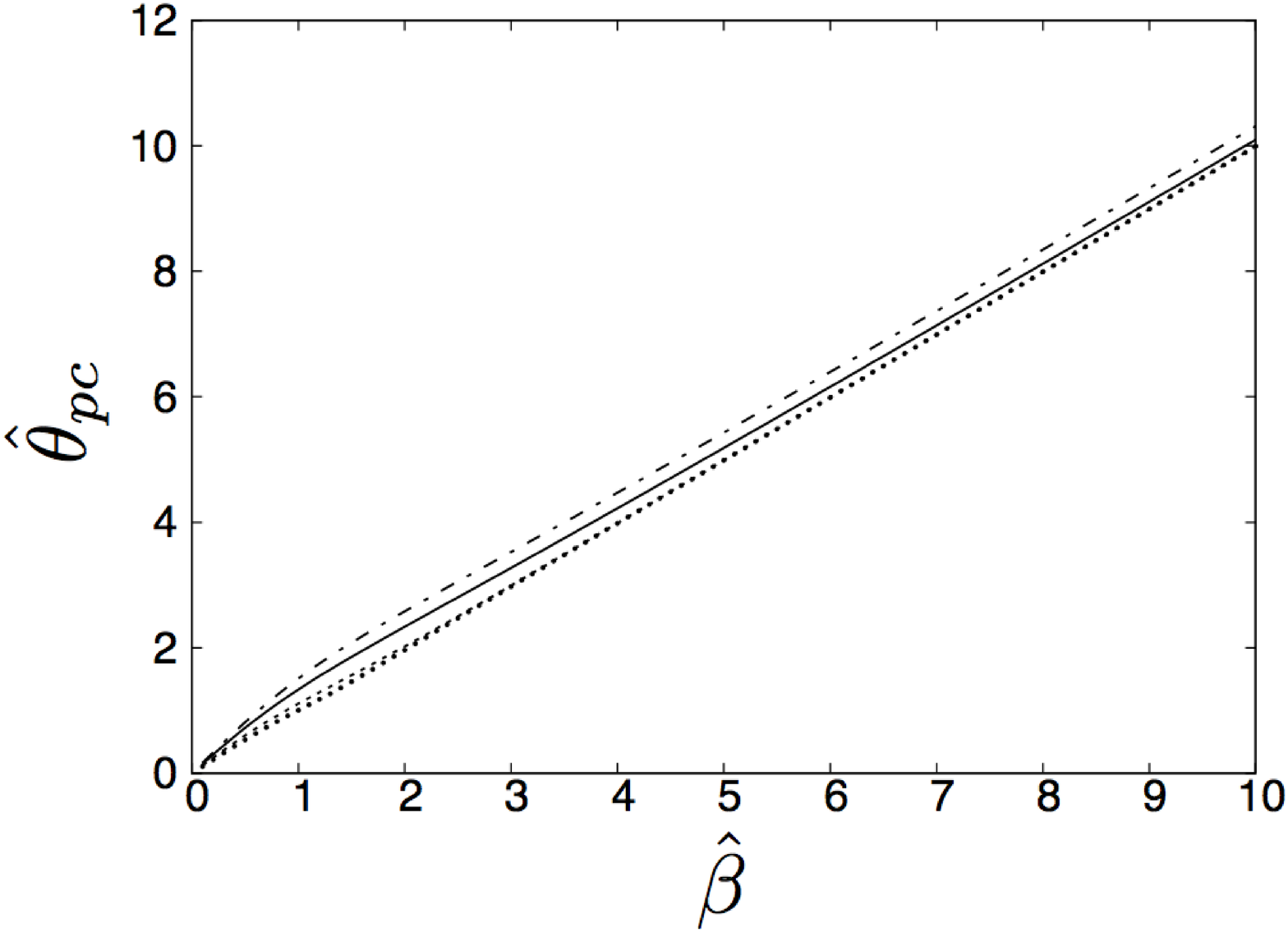}
\caption{
Image centroid $\hat\theta_{\it{pc}}$ and $\hat\beta$ 
for $\varepsilon > 0$ (convex-type attractive models). 
The dot-dashed, solid, dashed and dotted curves 
denote $n=0.5$, $1$, $3$ and $10$, respectively. 
The horizontal axis denotes the source position 
$\hat\beta$ normalized by the Einstein radius, 
and the vertical axis denotes $\hat\theta_{\it{pc}}$. 
}
\label{figure-4}
\end{figure}

\begin{figure}
\includegraphics[width=8cm]{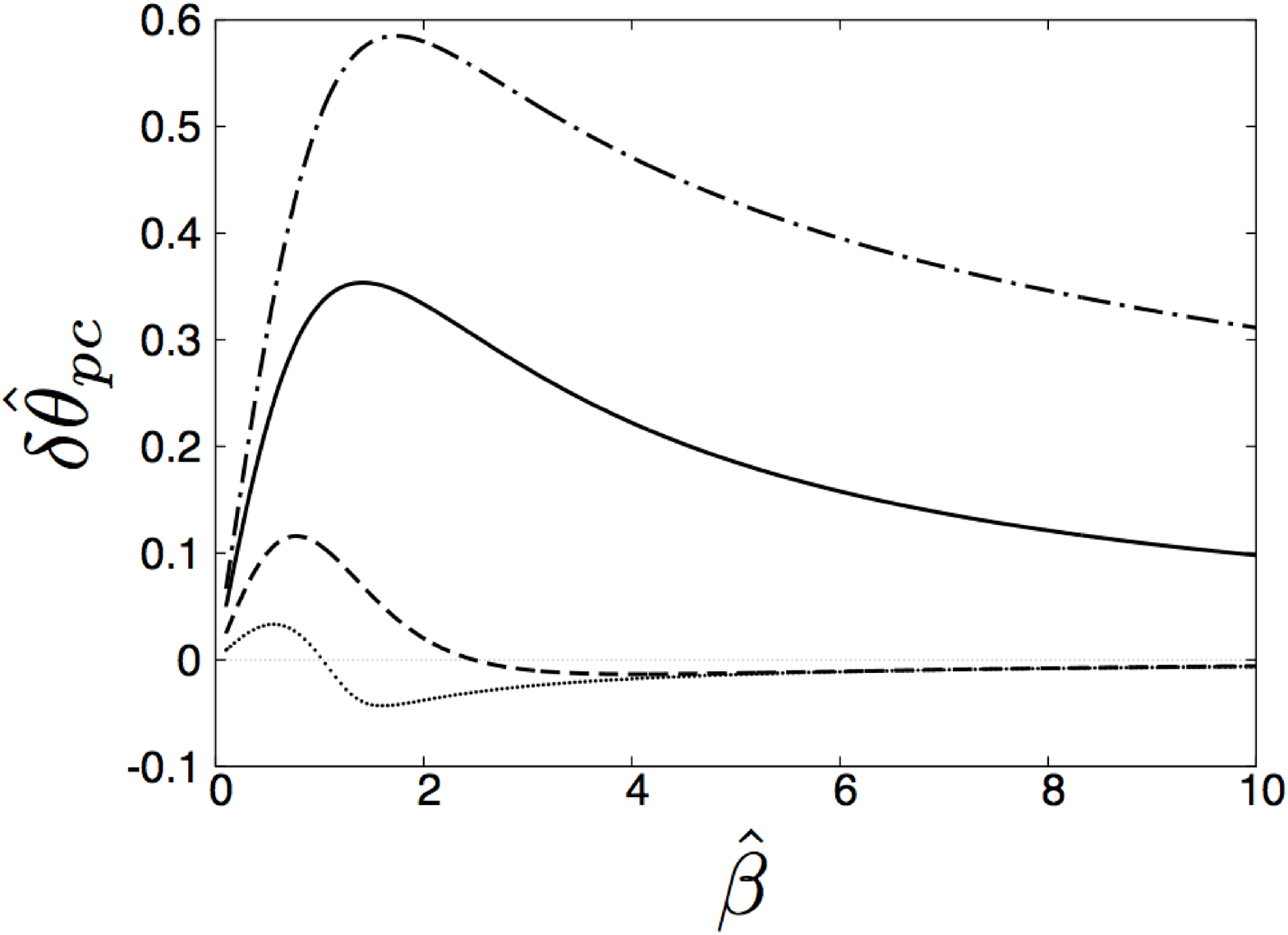}
\caption{
Image centroid shift $\delta\hat\theta_{\it{pc}}$ and $\hat\beta$ 
for $\varepsilon > 0$ (convex-type attractive models). 
The dot-dashed, solid, dashed and dotted curves 
denote $n=0.5$, $1$, $3$ and $10$, respectively. 
The horizontal axis denotes the source position 
$\hat\beta$ normalized by the Einstein radius, 
and the vertical axis denotes $\delta\hat\theta_{\it{pc}}$. 
}
\label{figure-5}
\end{figure}

At the center of the bow tie in the centroid shift, 
the image centroid position is the same as 
the intrinsic (unlensed) source position. 
At which time (and the corresponding source position) 
does the image centroid position agree with 
the source position? 
For Schwarzschild lens, the image centroid position 
agrees with the source position 
only at $t = \pm \infty$, 
namely $\beta = \infty$. 
In order to study this coincidence time (and source position), 
it is convenient to use Figure \ref{figure-4} 
for $\hat\theta_{\it{pc}}$ and $\hat\beta$ 
and Figure \ref{figure-5} 
for $\delta\hat\theta_{\it{pc}}$ and $\hat\beta$. 
Roughly speaking, the coincidence occurs 
at $\hat{\beta} \sim 1-3$, namely 
a few times the Einstein crossing time. 
This timescale might be used for 
applications to observations.

\subsection{Numerical computations: $\varepsilon < 0$ case} 
\begin{figure}
\includegraphics[width=8cm]{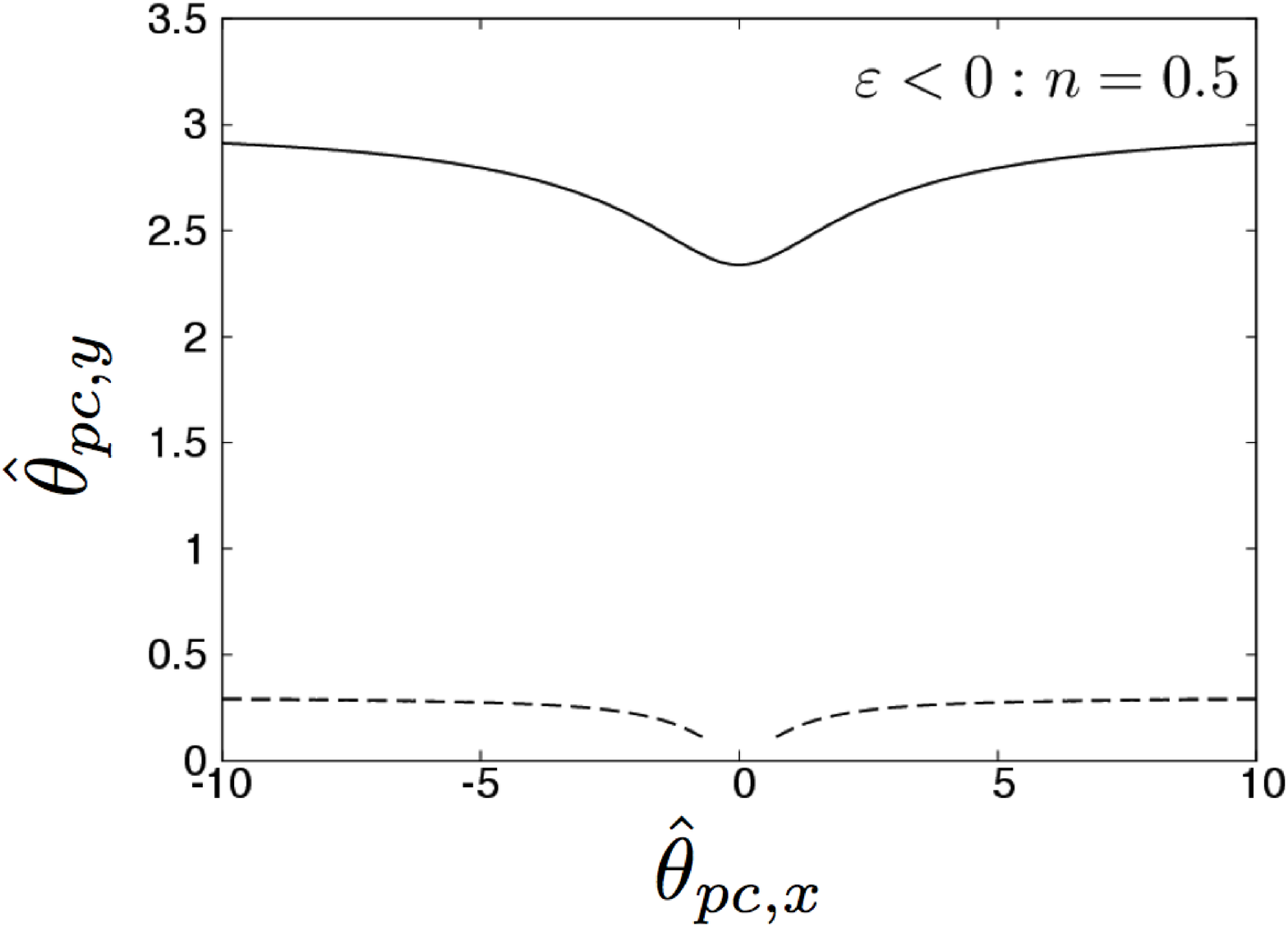}
\includegraphics[width=8cm]{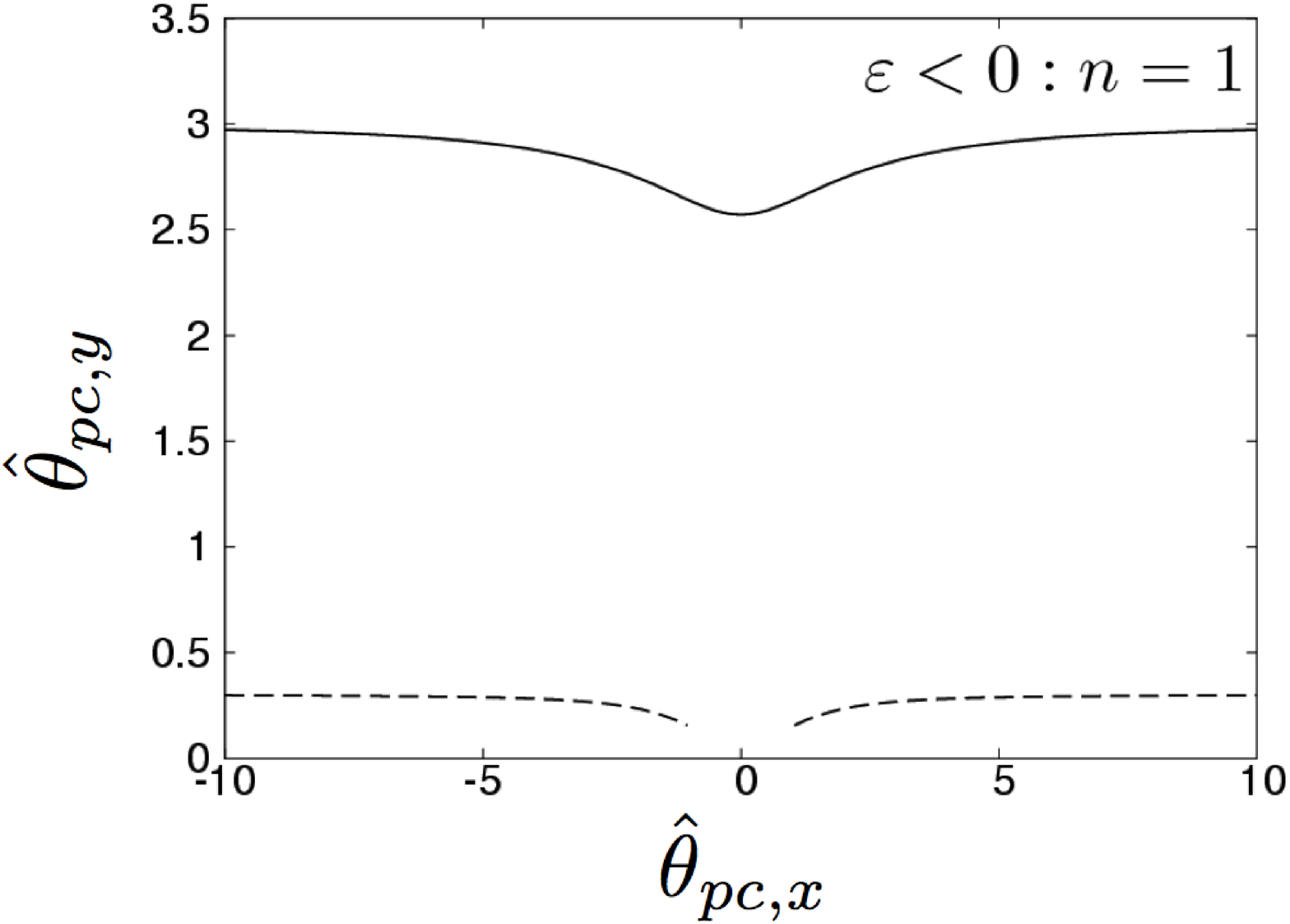}
\includegraphics[width=8cm]{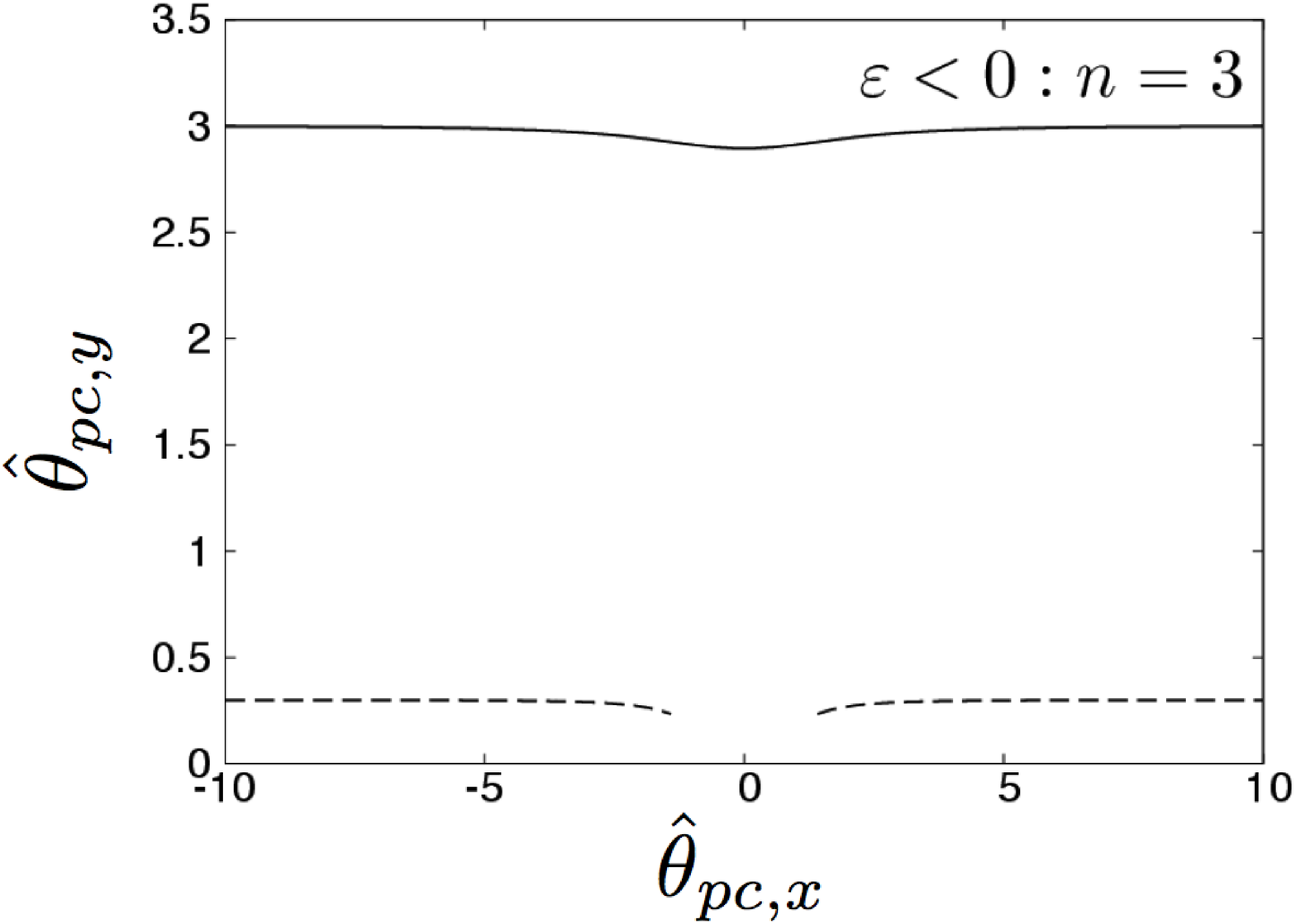}
\includegraphics[width=8cm]{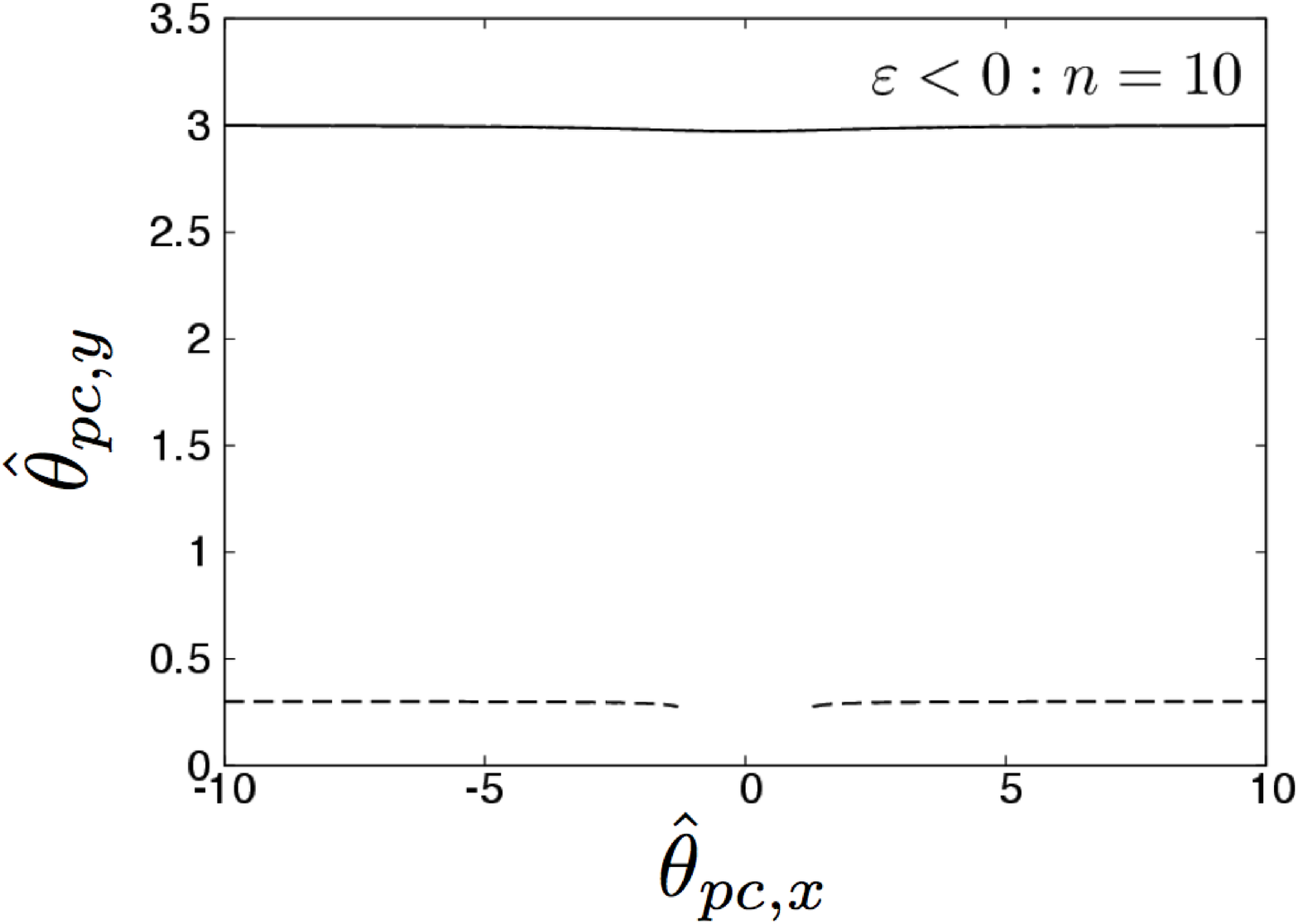}
\caption{
Centroid motions as $(\hat\theta_{\it{pc},x}, \hat\theta_{\it{pc},y})$ 
for $\varepsilon < 0$ (repulsive models). 
The solid and dashed curves 
correspond to $\hat\beta_0 = 3$ and $\hat\beta_0 = 0.3$, respectively. 
The horizontal axis along the source linear motion is $\hat\theta_{\it{pc},x}$ 
and the vertical axis is $\hat\theta_{\it{pc},y}$. 
The dashed curves do not exist for small $\hat\beta$, 
where no images appear. 
Top left: $n = 0.5$
Top right: $n = 1$. 
Bottom left: $n = 3$. 
Bottom right: $n = 10$. 
}
\label{figure-6}
\end{figure}

Next, we consider the $\varepsilon < 0$ case. 
Figure \ref{figure-6} shows the image centroid motion 
by the $\varepsilon < 0$ models. 
Note that the centroid curve does not exist for small $\hat\beta$ 
because of no images. 
See also Figure \ref{figure-1} for no image cases. 
Such a peculiar event might be misinterpreted 
as an eclipse in astronomy. 

\begin{figure}
\includegraphics[width=8cm]{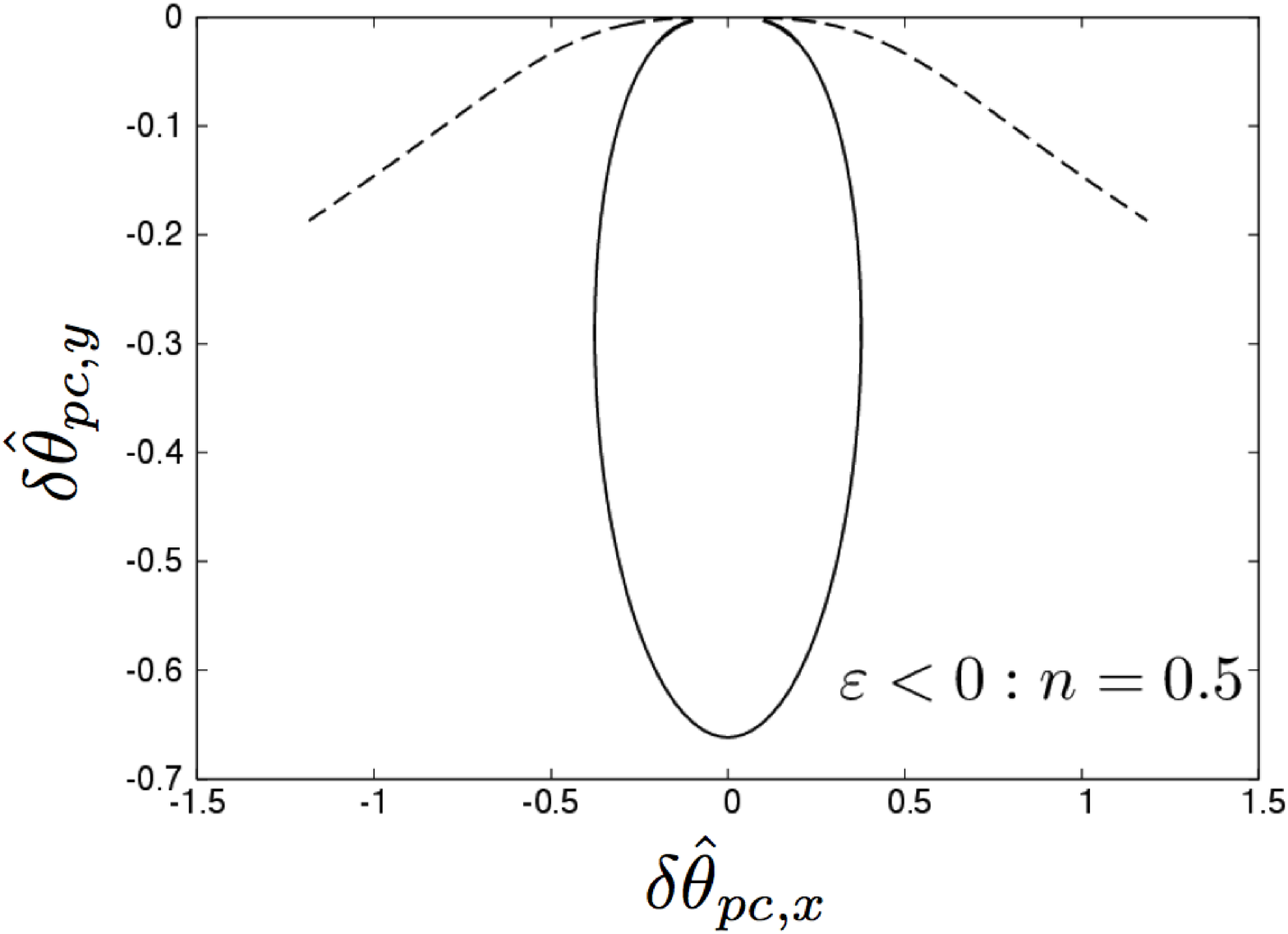}
\includegraphics[width=8cm]{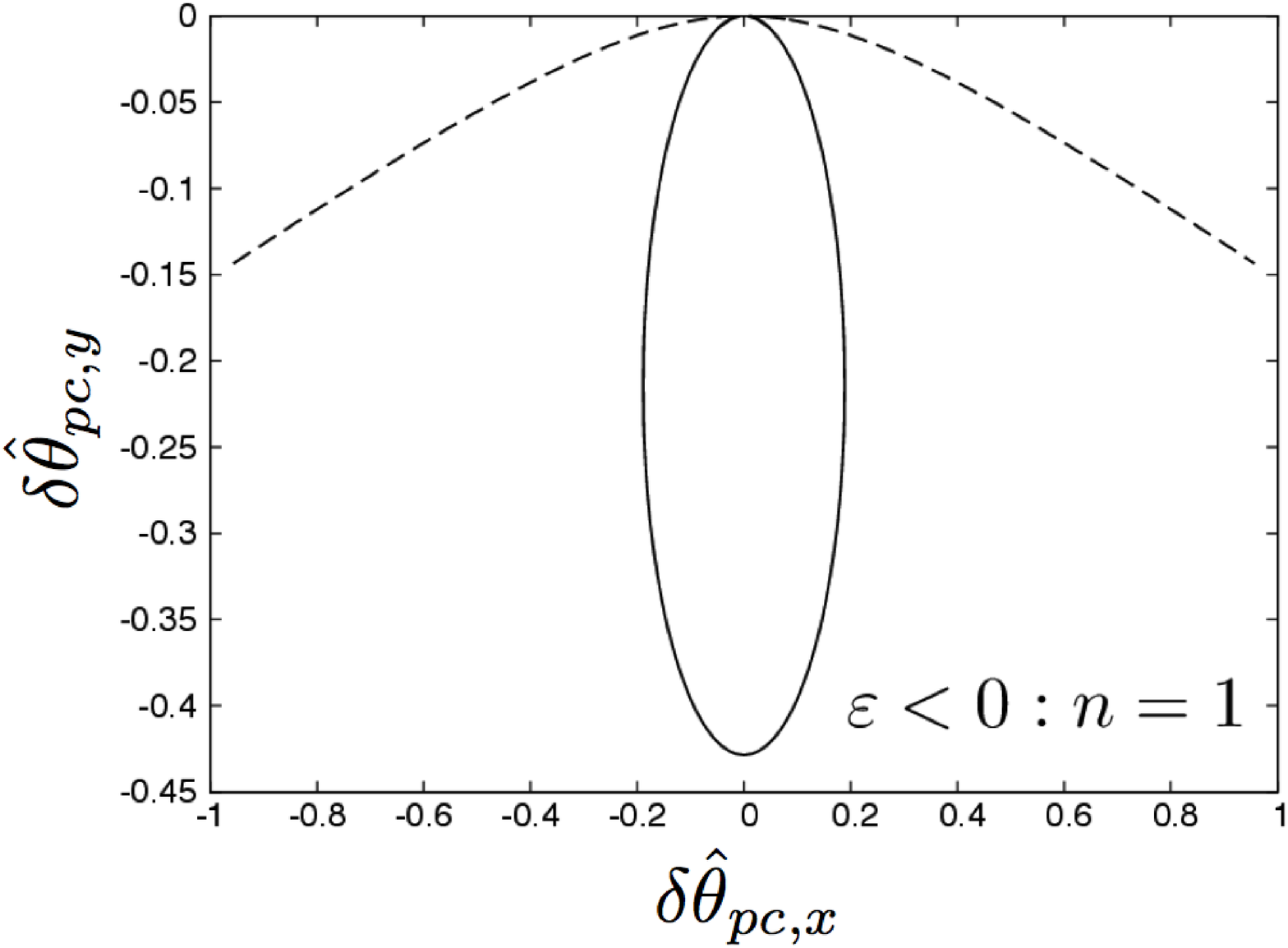}
\includegraphics[width=8cm]{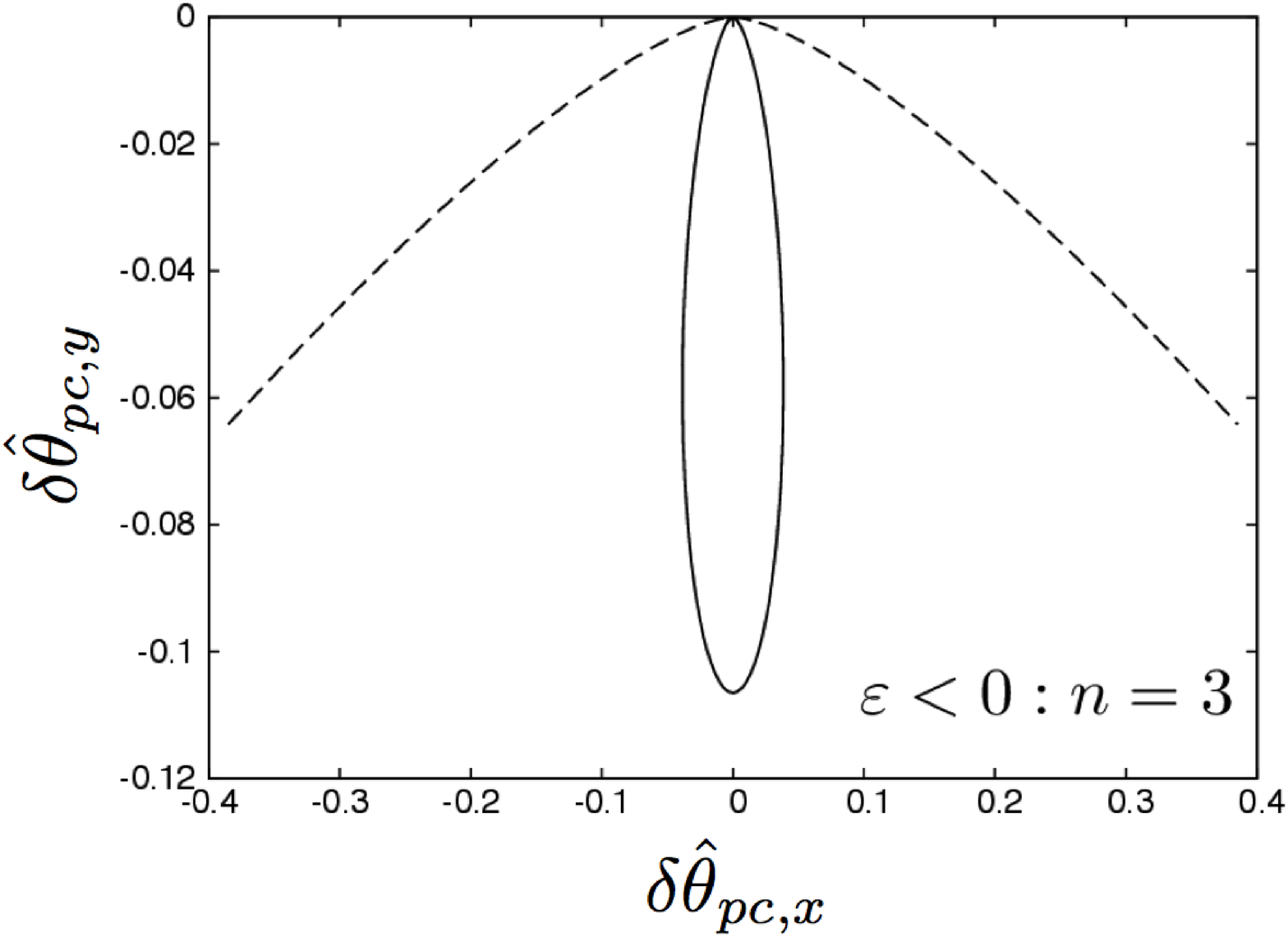}
\includegraphics[width=8cm]{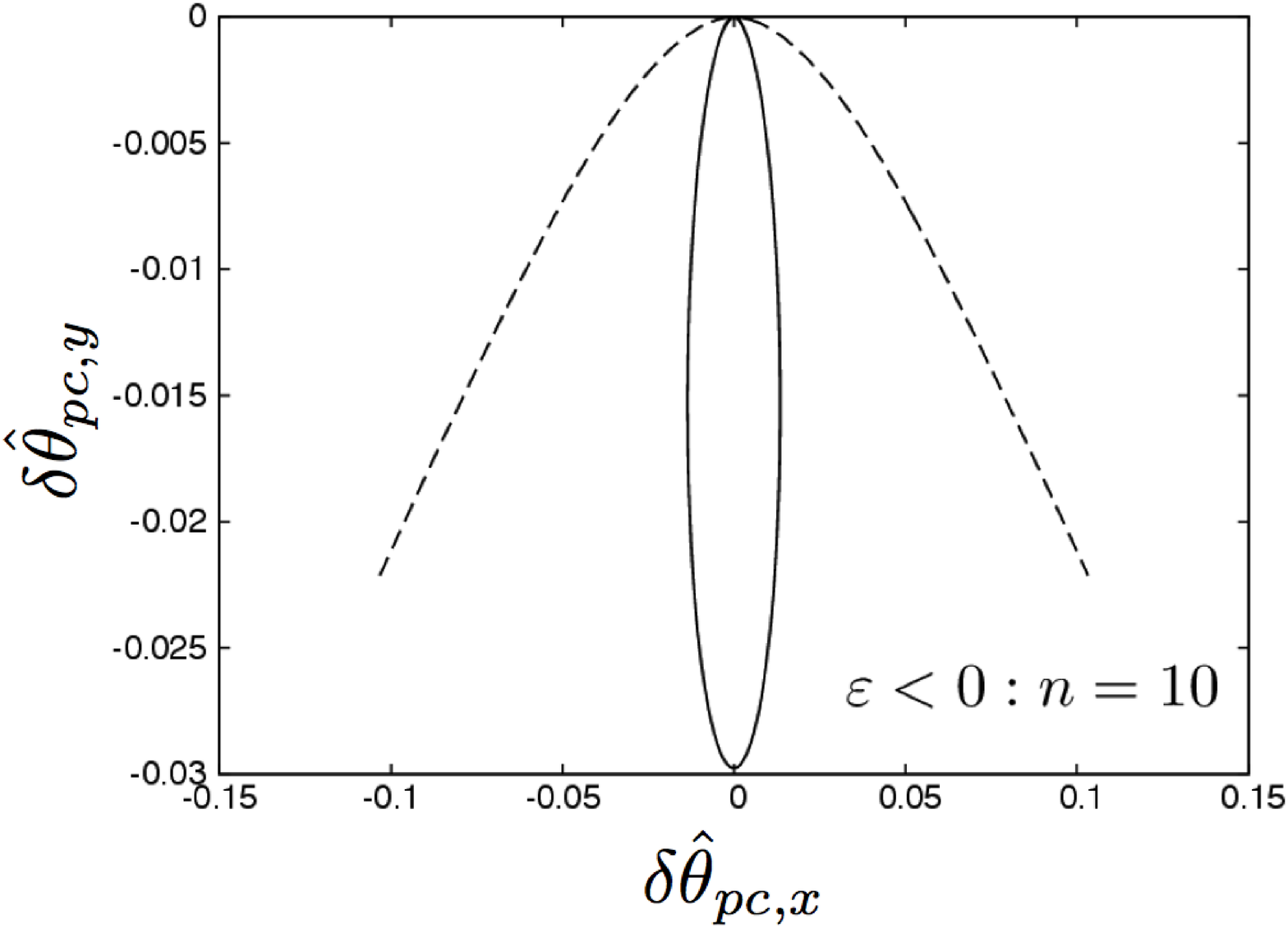}
\caption{
Centroid shifts $\delta\boldsymbol{\hat{\theta}}_{\it{pc}}$ 
for $\varepsilon < 0$ (concave-type repulsive models). 
The solid and dashed curves 
correspond to $\hat\beta_0 = 3$ and $\hat\beta_0 = 0.3$, respectively. 
The horizontal axis along the source velocity is $\delta\hat\theta_{\it{pc},x}$ 
and the vertical axis is $\delta\hat\theta_{\it{pc},y}$. 
The dashed curves are not closed, 
because no images appear for small $\hat\beta$. 
Top left: $n = 0.5$
Top right: $n = 1$. 
Bottom left: $n = 3$. 
Bottom right: $n = 10$. 
}
\label{figure-7}
\end{figure}

Figure \ref{figure-7} shows the image centroid shift 
by the $\varepsilon < 0$ models. 
There does not appear any bow-tie shape. 
Note that the image centroid shift is always negative, 
because the effective force is repulsive. 
For unseen lens objects, the negative shift can be hardly 
distinguished from the positive one. 

The centroid shift trajectory for the repulsive models 
might be elongated vertically to the source motion direction 
like a prolate spheroid as shown by Figure \ref{figure-7}, 
whereas that for convex-type attractive models 
such as the Schwarzschild one 
is tangentially elongated like an oblate spheroid 
(See Figure \ref{figure-3}). 
Figures $\ref{figure-3}$ and $\ref{figure-7}$ show that 
the size of the centroid shift by the repulsive models 
for each $n$ and $\hat\beta_0$ is comparable to 
that for the corresponding $\varepsilon > 0$ models.

\subsection{Parameter estimations} 
Equations (\ref{theta_E}) and (\ref{theta_E2}) 
are rewritten as 
\begin{eqnarray}
\frac{|\bar{\varepsilon}|}{R_{\it{E}}^n} 
&=& \cfrac{D_{S} R_{E}}{D_{\it{LS}} D_{\it{L}}} 
\nonumber\\
&=& \frac{D_{S} \theta_{E}}{D_{\it{LS}}} .
\label{observable}
\end{eqnarray}
Here, $D_{\it{L}}$, $D_{\it{S}}$, $D_{\it{LS}}$ and $R_{E} = D_{\it{L}} \theta_{\it{E}}$ 
are observables in astronomy, 
while $\bar{\varepsilon}$ and $n$ are not direct observables 
but model parameters. 
Note that $\bar{\varepsilon}/R_{E}^n$ is comparable to 
the typical size of the deflection angle. 

The right-hand side of Eq. (\ref{observable}) consists of 
the observables and it is a dimensionless quantity. 
Hence, Eq. (\ref{observable}) allows us to 
investigate $|\bar{\varepsilon}|/R_{E}^n$ from observations. 
See Tables \ref{table-2} and \ref{table-3} 
for Einstein ring size and Einstein radius crossing time, respectively. 
Near future astrometry space missions such as Gaia 
and JASMINE are expected to have angular sensitivity 
of a few micro arcseconds, for which the relevant parameter 
combination is limited as $|\bar{\varepsilon}|/R_{\it{E}}^n > 10^{-11}$. 
Roughly speaking, the mission life time is several years, 
for which the relevant timescale is limited as 
$t_{\it{E}} <$ a few years and Table \ref{table-3} thus 
tells the limit as $|\bar{\varepsilon}|/R_{\it{E}}^n < 10^{-7}$ (for Bulge) 
and $< 10^{-8}$ (for LMC). 
In total, the parameter range relevant for the near future missions 
is $10^{-11} < |\bar{\varepsilon}|/R_{\it{E}}^n < 10^{-7}$.

\begin{figure}
\includegraphics[width=12cm]{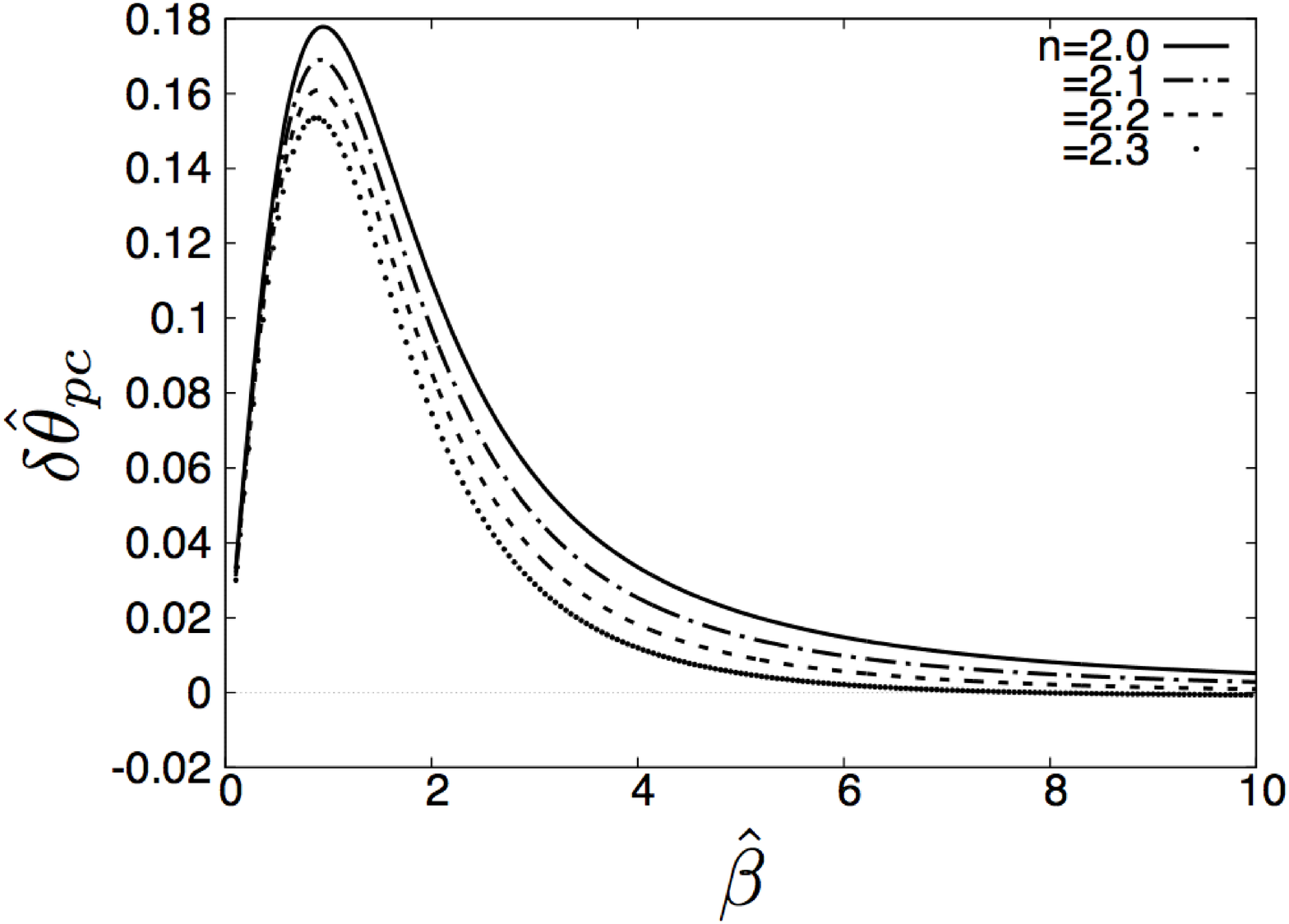}
\includegraphics[width=12cm]{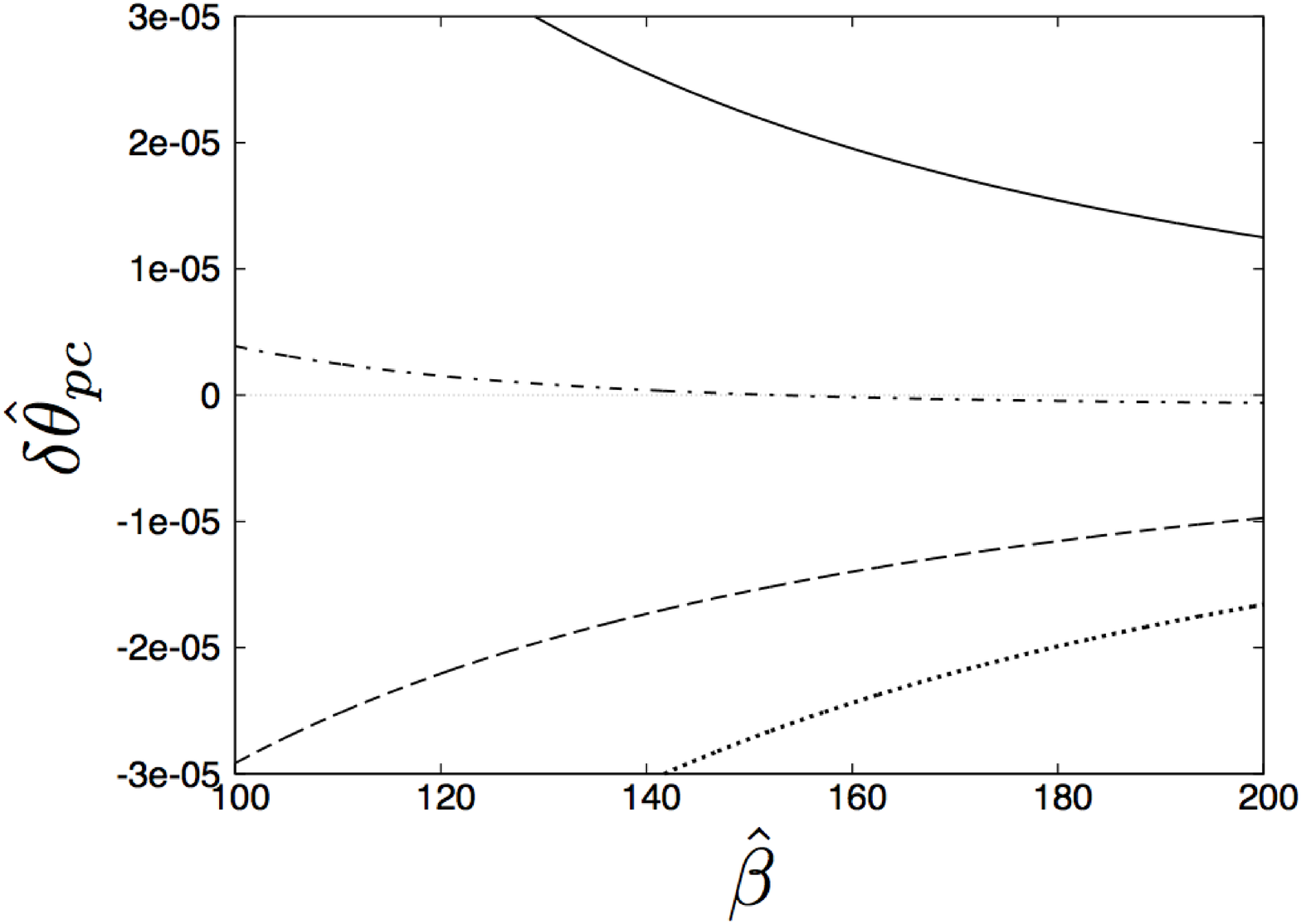}
\caption{
Image centroid shift $\delta\hat\theta_{\it{pc}}$ and $\hat\beta$ 
for $\varepsilon > 0$ (convex-type attractive models). 
The solid, dot-dashed, dashed and dotted curves 
denote $n=2.0$, $2.1$, $2.2$ and $2.3$, respectively. 
The horizontal axis denotes the source position 
$\hat\beta$ normalized by the Einstein radius, 
and the vertical axis denotes $\delta\hat\theta_{\it{pc}}$. 
Top: $\hat\beta \in [0, 10]$. 
Bottom: $\hat\beta \in [100, 200]$. 
}
\label{figure-8}
\end{figure}

Before closing this section, 
we mention how large $n$ models could lead to a multiply-connected 
curve of the microlensed centroid shift. 
Numerical calculations suggest that 
$n>2$ and $\varepsilon > 0$ models could produce 
a bow-tie shape. 
See also Figure \ref{figure-8} for numerical computations 
in the vicinity of $n=2$ as $n = 2.0, 2.1, 2.2$ and $2.3$. 
The numerical calculations suggest that 
the bow tie shape could appear, if $n > 2$. 
Numerical computations for other parameter values 
suggest that the maximum numbers of the loops and the knots 
in the the centroid curve are three and one, respectively, 
which are actually achieved by the $n=3$ model.

\section{Discussion and Conclusion}
We examined gravitational lens models 
inspired by modified gravity theories, exotic matter and energy. 
By using an asymptotically flat, static and spherically symmetric 
spacetime model of which metric depends on 
the inverse distance to the power of positive $n$, 
it was shown in the weak field and thin lens approximations that, 
for large $n$ cases in the convex-type models, 
the centroid shift from the source position 
might move on a multiply-connected curve like a bow tie, 
while it is known to move on an ellipse for $n=1$ case 
and to move on an oval curve for $n=2$. 
This bow-tie shape by the convex-type exotic lens models 
is distinguishable from standard ones due to binary motions 
or due the microlensing by Schwarzschild lens. 
The distinctive feature such as the bow-tie shape 
may be used for searching (or constraining) 
localized exotic matter or energy with astrometric observations. 

The parameter range relevant for the current and near-future missions 
such as Gaia and JASMIME 
is $10^{-11} < |\bar{\varepsilon}|/R_{\it{E}}^n < 10^{-7}$, 
where we assume that the accuracy in astrometry 
will reach a few micro arcseconds and the mission lifetime 
will be several years.

It was shown also that the centroid shift trajectory 
for concave-type repulsive models 
might be elongated vertically to the source motion direction 
like a prolate spheroid, 
whereas that for convex-type attractive models 
such as the Schwarzschild one 
is tangentially elongated like an oblate spheroid. 
The image centroid shift by the repulsive models is always negative, 
because the effective force is repulsive. 
For unseen lens objects, the negative shift can be hardly 
distinguished from the positive one. 
In this sense, it might be relatively difficult to investigate 
the repulsive models in astrometry.

We would like to thank F. Abe, M. Bartelmann, T. Harada, 
S. Hayward, J. Kunz, K. Nakao, Y. Sendouda, R. Takahashi, 
N. Tsukamoto, and M. Visser 
for the useful conversations on the exotic lens models.

\newpage

\begin{table}[h]
\caption{
The sign of the convergence $\kappa$. 
It is the same as that of $\varepsilon (1-n)$ 
according to Eq. (\ref{kappa}). 
}
  \begin{center}
    \begin{tabular}{c|c}
\hline
$\kappa > 0$  \; &  
\; 
$\varepsilon > 0 \; \& \; n <1$ \\
 \; & 
\; 
$\varepsilon < 0 \; \& \; n >1$ \\
\hline 
$\kappa = 0$ \; & 
$n =1$ \\
\hline
$\kappa < 0$ \; & 
\; 
$\varepsilon > 0 \; \& \; n >1$ \\
 \; & 
\; 
$\varepsilon < 0 \; \& \; n <1$ 
\\
\hline
    \end{tabular}
  \end{center}
\label{table-1}
\end{table}

\begin{center}
\begin{table}
\caption{Einstein radii and model parameters for Bulge and LMC lensings. 
$\theta_{\it{E}}$ is the angular Einstein radius, 
$R_{\it{E}}$ is the Einstein radius, and $\bar{\varepsilon}$ and $n$ are 
the two model parameters. 
$D_{\it{S}} = 8 kpc$ and $D_{\it{L}} = 4 kpc$ are assumed for Bulge. 
$D_{\it{S}} = 50 kpc$ and $D_{\it{L}} = 25 kpc$ are assumed for LMC. 
} 
\begin{tabular}{rrrrrrr}
\hline \hline
  & & \multicolumn{2}{c}{Bulge} & &  \multicolumn{2}{c}{LMC} \\ \cline{3-4} \cline{6-7}
  $\theta_{\it{E}} (mas)$ & &  $R_{\it{E}} (km)$ & $\frac{\bar{\varepsilon}}{R_{\it{E}}^n}$ & & $R_{\it{E}} (km)$ & $\frac{\bar{\varepsilon}}{R_{\it{E}}^n}$ \\
 \hline
$10^{-3}$  & & $6.0 \times 10^5$ & $1.0 \times 10 ^{-11}$ & & $3.7 \times 10^6$  & $1.0 \times 10 ^{-11}$ \\
$10^{-2}$  &  & $6.0 \times 10^6$ & $1.0 \times 10 ^{-10}$ & & $3.7 \times 10^7$ & $1.0 \times 10 ^{-10}$ \\
$10^{-1}$ &  & $6.0 \times 10^7$ & $1.0 \times 10 ^{-9}$ & & $3.7 \times 10^8$ & $1.0 \times 10 ^{-9}$ \\
$1$ &  & $6.0 \times 10^8$ & $1.0 \times 10 ^{-8}$ & & $3.7 \times 10^9$ & $1.0 \times 10 ^{-8}$ \\
$10$ &  & $6.0 \times 10^9$ & $1.0 \times 10 ^{-7}$ & & $3.7 \times 10^{10}$ &  $1.0 \times 10 ^{-7}$ \\
$10^2$ &  & $6.0 \times 10^{10}$ & $1.0 \times 10 ^{-6}$  & & $3.7 \times 10^{11}$ & $1.0 \times 10 ^{-6}$  \\
$10^3$ &  & $6.0 \times 10^{11}$ & $1.0 \times 10 ^{-5}$ & &  $3.7 \times 10^{12}$ & $1.0 \times 10 ^{-5}$  \\
\hline \hline
\end{tabular}
\label{table-2}
\end{table}
\end{center}

\begin{center}
\begin{table}
\caption{Einstein radius crossing times for Bulge and LMC lensings. 
$t_{\it{E}}$ is the Einstein radius crossing time. 
$D_{\it{S}} = 8 kpc$ and $D_{\it{L}} = 4 kpc$ are assumed for Bulge. 
$D_{\it{S}} = 50 kpc$ and $D_{\it{L}} = 25 kpc$ are assumed for LMC. 
$v_T = 220 km/s$ is assumed for Bulge and LMC.
In this table, the Einstein radius is calculated by 
$R_{\it{E}} = v_T \times t_{\it{E}}$ from the definition of 
the Einstein radius crossing time. 
Here, the input is $t_{\it{E}} \sim 10^{-3} - 10^3 (day)$, 
namely $1 (min.) - 3 (yr.)$.  
}
\begin{tabular}{rrrrrrrr}
\hline \hline
$t_{\it{E}} (day)$ & & $R_{\it{E}} (km)$ & & $\frac{\bar{\varepsilon}}{R_{\it{E}}^n}$ [Bulge] 
& & $\frac{\bar{\varepsilon}}{R_{\it{E}}^n}$ [LMC] \\
\hline
$10^{-3}$ & & $1.9 \times 10^4$ & & $3.1 \times 10^{-13}$ 
& & $5.0 \times 10^{-14} $ \\
$10^{-2}$ & & $1.9 \times 10^5$ & & $3.1 \times 10^{-12}$ 
& & $5.0 \times 10^{-13} $ \\
$10^{-1}$ & & $1.9 \times 10^6$ & & $3.1 \times 10^{-11}$ 
& & $5.0 \times 10^{-12} $ \\
$1$ & & $1.9 \times 10^7$ & & $3.1 \times 10^{-10}$ 
& & $5.0 \times 10^{-11} $ \\
$10$ & & $1.9 \times 10^8$ & & $3.1 \times 10^{-9}$ 
& & $5.0 \times 10^{-10} $ \\
$10^{2}$ & & $1.9 \times 10^9$ & & $3.1 \times 10^{-8}$ 
& & $5.0 \times 10^{-9} $ \\
$10^{3}$ & & $1.9 \times 10^{10}$ & & $3.1 \times 10^{-7}$ 
& & $5.0 \times 10^{-8} $ \\
\hline \hline
\end{tabular}
\label{table-3}
\end{table}
\end{center}


\begin{thebibliography}{99}
\bibitem{Note-1}
Here, we mention the truncated expression in Eq. (\ref{ds}). 
Let us begin with considering a sphere whose radius from 
the center is denoted as $r$. 
The mass inside the sphere is denoted as $M(r)$. 
Any physically reasonable theory of gravity must admit 
the Newtonian limit, 
in accordance with experiments and observations.
Hence, the metric components $g_{tt}$ and $g_{rr}$ 
in the weak field region are expected to 
have the Newton-type potential written as $M(r) / r$. 
Note that the part of $1/r$ comes from the Green function for 
the Laplacian operator (and thus the Poisson equation). 
The fraction as $M(r) / r$ becomes the inverse distance 
only for a point mass case, while it is a more general function 
of the distance for extended objects. 
Roughly speaking, therefore, the ansatz of the space-time metric 
Eq. (\ref{ds}) might correspond to assuming $M(r)$ 
in the power law of $r$. 
For example,  $n=0$, $n=1$ and $n=2$ correspond to 
isothermal spheres (as a simple model of a galaxy), 
the vacuum region outside of an bounded mass distribution, 
and Ellis worm holes, respectively. 
Ellis worm holes are non-vacuum and non-singular (regular) 
in the sense that they are associated with 
a particular scalar field configuration. 
Models under study thus describe a regular distribution of matter 
for $n \neq 1$.
\end{thebibliography}
\end{document}